\documentclass[%
 reprint,
superscriptaddress,
nofootinbib,
 amsmath,amssymb,
 aps,
]{revtex4-1}

\usepackage[dvipsnames]{xcolor}
\usepackage{tikz}
\usepackage{graphicx}
\usepackage{dcolumn}
\usepackage{bm}
\usepackage{hyperref}
\usepackage{enumitem}
\usepackage{lineno}

\usepackage{multirow}
\newcommand\Tstrut{\rule{0pt}{3.3ex}}         
\newcommand\Bstrut{\rule[-2ex]{0pt}{0pt}}   

\hypersetup{
  colorlinks=true,
  citecolor=blue,
  urlcolor=purple
}
\usepackage{aas_macros}

\newcommand{\B}[1]{\ensuremath{\boldsymbol{#1}}}

\newcommand{\iMpc}{\ensuremath{h/\mathrm{Mpc}}}

\newcommand{\Mpc}{\ensuremath{\mathrm{Mpc}/h}}
\newcommand{\Gpc}{\ensuremath{\mathrm{Gpc}/h}}

\newcommand{\dittotikz}{%
    \tikz{
        \draw [line width=0.12ex] (-0.2ex,0) -- +(0,0.8ex)
            (0.2ex,0) -- +(0,0.8ex);
        \draw [line width=0.08ex] (-0.6ex,0.4ex) -- +(-0.0em,0)
            (0.6ex,0.4ex) -- +(0.0em,0);
    }%
}

\def\fun#1#2{\lower3.6pt\vbox{\baselineskip0pt\lineskip.9pt
        \ialign{$\mathsurround=0pt#1\hfill##\hfil$\crcr#2\crcr\sim\crcr}}}

\newcommand{\beq}{\begin{equation}}
\newcommand{\eeq}{\end{equation}}
\newcommand{\beqa}{\begin{eqnarray}}
\newcommand{\eeqa}{\end{eqnarray}}

\newcommand{\correction}[1]{\textcolor{Black}{#1}}

\begin{document}


\title{Testing one-loop galaxy bias: Power spectrum}


\author{Alexander Eggemeier}
\email{alexander.eggemeier@durham.ac.uk}
\affiliation{%
  Institute for Computational Cosmology, Department of Physics, Durham University, South Road, Durham DH1 3LE,
  United Kingdom
}%

\author{Rom\'an Scoccimarro}
\affiliation{
  Center for Cosmology and Particle Physics, Department of Physics, New York University, NY 10003, New York, USA
}%

\author{Martin Crocce}
\affiliation{%
  Institute of Space Sciences (ICE, CSIC), Campus UAB, Carrer de Can Magrans, s/n, 08193 Barcelona, Spain
}%
\affiliation{%
  Institut d’Estudis Espacials de Catalunya (IEEC), 08034 Barcelona, Spain
}%

\author{Andrea Pezzotta}
\affiliation{%
  Institute of Space Sciences (ICE, CSIC), Campus UAB, Carrer de Can Magrans, s/n, 08193 Barcelona, Spain
}%
\affiliation{%
  Institut d’Estudis Espacials de Catalunya (IEEC), 08034 Barcelona, Spain
}%
\affiliation{%
  Max-Planck-Institut f\"{u}r extraterrestrische Physik, Postfach 1312, Giessenbachstr., 85741 Garching, Germany
}%

\author{Ariel G. S\'{a}nchez}
\affiliation{%
  Max-Planck-Institut f\"{u}r extraterrestrische Physik, Postfach 1312, Giessenbachstr., 85741 Garching, Germany
}%

\date{\today}

\begin{abstract} 
  We test the regime of validity of one-loop galaxy bias for a wide variety of biased tracers.  Our most
  stringent test asks the bias model to simultaneously match the galaxy-galaxy and galaxy-mass spectrum, using
  the measured nonlinear matter spectrum from the simulations to test one-loop effects from the bias expansion
  alone.  In addition, we investigate the relevance of short-range nonlocality and halo exclusion through
  higher-derivative and scale-dependent noise terms, as well as the impact of using co-evolution relations to
  reduce the number of free fitting parameters. From comparing validity and merit of these assumptions we find
  that a four-parameter model (linear, quadratic, cubic nonlocal bias, and constant shot noise) with fixed
  quadratic tidal bias provides a robust modeling choice for the auto power spectrum of the less massive halos
  in our set of samples and their galaxy populations (up to $k_{\mathrm{max}} = 0.35\,\iMpc$ for a sample volume
  of $6\,(\Gpc)^3$). For the more biased tracers it is most beneficial to include scale-dependent noise. This is
  also the preferred option when considering combinations of the auto and cross power spectrum, which might be
  relevant in joint studies of galaxy clustering and weak lensing. We also test the use of perturbation theory
  to account for matter loops through gRPT, EFT and the hybrid approach RESPRESSO. While all these have similar
  performance, we find the latter to be the best in terms of validity and recovered mean posterior values, in
  accordance with it being based partially on simulations. \newline\newline\newline
\end{abstract}

\pacs{Valid PACS appear here}
\maketitle


\section{Introduction}
\label{sec:introduction}

The varying degrees of clustering displayed by different types of galaxies, or clusters of galaxies, has led to
the understanding that these objects cannot be unbiased tracers of the underlying matter distribution
(e.g. \cite{HauPee7311,Kaiser:1984,DavEfsFre8505,BarBonKai8605,TegBlaStr04,ZehZheWei05}, for a recent review see 
\cite{Desjacques:2018}). In order to utilize measurements from large-scale structure surveys for cosmological
studies, it is therefore of critical importance to have a robust model of the galaxy-matter connection, commonly
referred to as \emph{galaxy bias}. The accuracy and consistency of these models will be challenged by great
improvements in the statistical precision of upcoming survey generations, e.g. DESI \cite{DESI} and Euclid
\cite{Euclid}, as well as the combination of multiple probes, such as clustering and weak lensing. In this light
it is interesting to consider the range of scales over which we can trust our currently best descriptions of
galaxy bias and how much freedom we need to allow for when analyzing the two-point clustering of \mbox{galaxies
  --- both} largely unresolved questions.

We intend to address these questions in the context of the perturbative galaxy bias expansion, which relates the
galaxy density contrast $\delta_g$ to a series of physically motivated terms that involve the matter density
fluctuations $\delta$ and its tidal field. The well-known linear relationship $\delta_g = b_1\,\delta$
\cite{Kaiser:1984}, with linear bias parameter $b_1$, represents the lowest order term in this expansion and is
valid only on the largest scales. Additional terms with their own bias parameters become relevant on
successively smaller scales, such as powers of the matter field \cite{Coles:1993,FryGaz9308}, $b_2\,\delta^2$,
$b_3\,\delta^3$ etc., as expected from a spherically-symmetric gravitational collapse
\cite{MoWhi9609,MoJinWhi97}. Similarly it has been argued that anisotropies in this process should lead to a
dependence on the tidal field \cite{CatLucMat9807,Catelan:2000,McDonald:2009,Matsubara:2011}, which was
confirmed by the inadequacy of the power series expansion to fully explain the clustering of dark matter
halos \cite{ManGaz1107,RotPor1107} and inconsistencies between two- and three-point statistics
\cite{Pollack:2012,PolSmiPor1405}. The first direct evidence from simulated data was reported in
\cite{Chan:2012,Baldauf:2012}. Further developments have put these arguments on a more solid theoretical footing
by identifying the matter density and tidal field as the leading, local gravitational effects that leave an
imprint on galaxy formation as a result of the equivalence principle \cite{Sen1406,DaiPajSch1511}. In addition,
since the evolution of galaxies occurs over long timescales, one should not expect the galaxy density to depend
on these quantities at only a single point in time, but rather on their entire past lightcone. It has been shown
that this time dependence can be traded for a set of extra (nonlocal) terms at each order of perturbation theory that are
generated by time evolution \cite{McDonald:2009,Chan:2012}. Systematic procedures to identify these terms have
been presented in \cite{Mirbabayi:2015,Desjacques:2018,EggScoSmi1906} and provide a complete basis at fixed time
for the general bias expansion.

An important assumption that goes into the derivation of this basis is that galaxy formation is spatially
local. It is well known that galaxies collect matter from an extended region of space, but as this is roughly
limited to the Lagrangian radii $R$ of their host halos, the spatially local assumption must be valid in the
large-scale limit. However, corrections known as \emph{higher-derivative} terms can become relevant on scales
approaching $R$, starting with a term $R^2\,\nabla^2\delta$
\cite{Des0811,McDonald:2009,DesCroSco1011}. Furthermore, whether a galaxy forms at a given point will not be
solely determined by the large-scale fields, but also have a dependency on the very small scale modes. In the
absence of strong primordial non-Gaussianities they are uncorrelated with any large-scale effects and so, from
the perturbative point of view, look like a stochastic contribution \cite{Dekel:1999,TarSod9909,Mat9911}. This
gives rise to a constant offset from Poisson shot noise in the galaxy power spectrum \cite{Scherrer:1998}, which
can be interpreted as being due to the halo exclusion effect \cite{MoWhi9609,SheLem99,SmiScoShe0703}. On smaller
scales halo-halo exclusion imprints a scale-dependence on the stochasticity, whose strength is controlled by the
Lagrangian radius \cite{BalSelSmi1310,Desjacques:2018} like for the higher-derivative terms. Up to date it is
not clear which of the two, if any, might have the more dominant effect on galaxy clustering.

Several tests of this bias modeling, to varying levels of detail and either in configuration or Fourier space,
have already been conducted in the literature. One of these studies \cite{SaiBalVla1405} applied the model to
the cross power spectrum and bispectrum between the dark matter field and halos in various mass bins and at
multiple redshifts. Across these various samples they found good agreement with their simulation measurements
and consistency in the fitted bias parameters from the two statistics up to scales of $k \sim
0.1\,\iMpc$. Furthermore, they demonstrated the need for a cubic nonlocal bias contribution and
presented indications that its associated parameter as well as the quadratic one corresponding to the tidal field closely
follow the so-called local-Lagrangian (LL) relations (see also~\cite{BiaDesKeh1405}).

These relations arise by making the assumption that at
some time in the far past the galaxy overdensity can be exhaustively described by powers of $\delta$ alone. Upon
translation to later times, all remaining terms identified as part of the general bias expansion appear, but
their amplitudes are fixed in terms of $b_1$, $b_2$ etc. and so, if this assumption is merited, the LL relations
provide a valuable reduction of the modeling degrees of freedom.

An analysis similar to \cite{SaiBalVla1405} has
been presented in \cite{AngFasSen1509} including the halo auto power spectrum and a higher-derivative parameter
that was not taken into account by \cite{SaiBalVla1405}. Using the goodness-of-fit as an indicator, they
reported an accurate match to the data up to scales of $k = 0.3\,\iMpc$. Building on this work, the authors of
\cite{FujMauSen2001} also considered the possibility of scale-dependent stochasticity, but were neither able to
claim a clear detection of such an effect nor a contribution from the higher-derivative term. Moreover, none of
these studies assessed how many free model parameters are actually necessary to describe their measurements
\mbox{--- a} subject that was addressed in \cite{FonRegSee1805,WerPor2002} by means of a Bayesian model
selection criterion. From fitting the auto halo power spectrum in redshift space, \cite{FonRegSee1805} thus
found a slight preference for the LL model over leaving the corresponding bias parameters free, while
 \cite{WerPor2002} came to the conclusion that a four parameter model including $b_1$, $b_2$, as well as the
tidal and higher-derivatives biases, performs best for their real-space halo cross power spectrum (without
having considered application of the LL relations) and scales up to $k = 0.2\,\iMpc$. However, neither of these
two analyses determined clearly over which ranges of scales their preferred models remain valid and both
considered only a single sample of biased tracers.

The theoretical developments implemented in these simulation studies have also been successfully applied to
clustering measurements from the BOSS galaxy survey~\cite{SanScoCro1701,GilPerVer1702,BeuSeoSai1704,GriSanSal1705,AlaAtaBai1709} and more
recently in \cite{IvaSimZal1909,AmiGleKok1909,TroSanAsg2001}. The exact modeling choices differ in all these
cases, e.g. \cite{BeuSeoSai1704} employs the LL relations for both the tidal second-order bias and the nonlocal third-order 
bias parameter, whereas only the former is fixed in this way by~\cite{SanScoCro1701,GriSanSal1705,TroSanAsg2001}
and~\cite{IvaSimZal1909} sets the latter to zero, but leaves the tidal bias parameter free. Each of these
analyses involved a considerable effort in validating their models, consisting of fits using large sets of mock
catalogs and blinded mock challenges (e.g. \cite{NisAmiIva2003}) in order to carefully test whether they obtain
unbiased results in their cosmological parameters. However, there is little variety in the galaxy samples used
in these tests as they are tuned to reproduce the clustering properties of the observed galaxies, and beyond a
determination of the range of validity of the respective fiducial model no further systematic checks of the
various assumptions are carried out. The fact that they still arrive at comparable cosmological constraints
might mean that the differences in the galaxy bias modeling are negligible compared to the statistical
uncertainties in the BOSS survey, but this is likely going to change with DESI and Euclid.

Given the large variety of results in the literature, our goal in this paper is to present a rigorous and
systematic approach to testing the bias modeling that combines the strengths of several previous studies. We
take a large pool of different tracers --- three galaxy samples resembling the SDSS main galaxies, and the LOWZ
and CMASS samples of BOSS, as well as four halo catalogs in different mass bins and redshifts --- and analyze
their two-point clustering with identical models and parameter priors using the same range of scales and with
statistical uncertainties all corresponding to an \emph{effective} volume of $6\,(\Gpc)^3$. We want to conduct a
precise test of galaxy bias alone and for that reason do not take into account redshift space distortions. This
would introduce further modeling uncertainties, whose potential deficiencies could be absorbed by the bias
parameters, or vice versa. Moreover, we choose to keep the cosmological parameters fixed, which allows us to
decouple the modeling of the bias contributions from the nonlinear matter power spectrum by replacing the latter
with its simulation measurement. In this framework we evaluate the range of validity of different assumptions,
such as the application of LL relations and/or inclusion of higher-derivative and scale-dependent stochasticity,
from a combination of two performance metrics: 1) the unbiased-ness of recovered parameters, and 2) the
goodness-of-fit. Since models with more degrees of freedom give rise to larger parameter uncertainties, we
compare these validity estimates with a measure of the model's merit to discern how many and which parameters
constitute an optimal choice, following~\cite{OsaNisBer1903}. We perform this analysis not only for the
auto power spectrum, but also in combination with the cross power spectrum between galaxies/halos and matter, as
consistency between the two statistics provides an even more stringent test of the bias model and is of great
relevance for joint studies of galaxy clustering and weak lensing data.

Our paper is organized as follows. In Section~\ref{sec:theory} we start off with an overview of the galaxy power
spectrum at next-to-leading order in perturbation theory including the various contributions mentioned above. We
then present our simulated data sets in Section~\ref{sec:data}, along with the measurements, our fitting procedure
and prior choices, as well as a precise definition of the performance metrics employed in this
work. Section~\ref{sec:results_true_matter} reports our main findings on the model's range of validity and a
determination of the importance of higher-derivative and scale-dependent stochastic terms using the measured
nonlinear matter power spectrum. In Section~\ref{sec:results_models} we explore how these findings are affected when the
latter is calculated from different approaches using perturbation theory. Finally, we give our
conclusions in Section~\ref{sec:conclusions}.

\section{One-loop perturbation theory for biased tracers}
\label{sec:theory}

Recent analyses of galaxy surveys typically rely on one-loop perturbation theory predictions to model the
two-point statistics of biased tracers. The following sections serve to provide an overview of the relevant
results and various employed modeling assumptions, as well as to establish our notation to be used throughout
the remainder of this paper.

\subsection{Galaxy bias expansion}
\label{sec:galaxy-bias}

One-loop corrections to the linear theory predictions for two-point statistics arise from terms up to third
order in perturbations. For our purposes it is therefore sufficient to write the galaxy bias expansion as
(following \cite{EggScoSmi1906}, which also contains relations to other popular bias parametrizations)
\begin{equation}
  \label{eq:theory.dg_expansion}
  \begin{split}
    \delta_g(\B{x}) = \;\, &\bar{b}_1\,\delta(\B{x}) + \bar{\beta}_1\nabla^2\delta(\B{x}) +
    \bar{b}_2\,\delta^2(\B{x}) + \bar{\gamma}_2\,{\cal G}_2(\Phi_v|\,\B{x}) \\ &+ 
    \bar{\gamma}_{21}\,{\cal G}_2(\varphi_2,\varphi_1|\,\B{x}) + \ldots\,,    
  \end{split}
\end{equation}
where, for brevity, we have omitted all dependencies on redshift. Since $\delta_g$ is a scalar, the matter tidal field first enters at second order
of this expansion described by the \emph{Galileon}
\begin{equation}
  \label{eq:theory.G2def}
  {\cal G}_2(\Phi_v) \equiv \left(\nabla_{ij} \Phi_v\right)^2 - \left(\nabla^2\Phi_v\right)^2\,.
\end{equation}
It is expressed here in terms of second
derivatives of the normalized velocity potential $\Phi_v$, which is linked to the divergence of the matter
velocity field $\theta$ via \mbox{$\nabla^2\Phi_v = \theta$}. The nonlocal nature of gravitational
collapse requires the existence of further terms in the bias expansion, the first of which appears at third
order and is given by \cite{Chan:2012}
\begin{equation}
  \label{eq:theory.G21def}
  {\cal G}_2(\varphi_2,\varphi_1) \equiv \nabla_{ij}\varphi_2\,\nabla_{ij}\varphi_1 -
  \nabla^2\varphi_2\,\nabla^2\varphi_1\,,
\end{equation}
where $\varphi_1$ is the linear Lagrangian perturbation theory (Zel'dovich approximation) potential, satisfying the Poisson equation
$\nabla^2\varphi_1 = -\delta$. The nonlocality of gravitational evolution manifests in all higher-order corrections to the
Zel'dovich approximation~\cite{KofPog9503} starting from the second-order potential, defined by
$\nabla^2\varphi_2 = -{\cal G}_2(\varphi_1)$.

In addition to these terms, we have also included the leading higher-derivative contribution $\nabla^2\delta$ in
Eq.~(\ref{eq:theory.dg_expansion}). Such terms are expected as the formation of dark matter halos and galaxies
necessarily depends on the initial matter distribution within some finite region, which leads to a short-range
nonlocality \cite{BarBonKai8605,Mat9911,McDonald:2009,DesCroSco1011,MusShe1206} that needs to be distinguished
from that due to gravitational evolution alone. They are therefore tied to a particular length scale, which for
halos is close to their Lagrangian radius \cite{McDonald:2009, LazSch1911}, but can potentially be significantly
altered due to baryonic physics for certain types of observable tracers \cite{Desjacques:2018}. The existence of
such a length scale implies a crucial difference between the higher-derivative contributions and the remaining
terms in Eq.~(\ref{eq:theory.dg_expansion}), whose relative importance is entirely governed by the nonlinearity
scale of the matter density.  If we assume here that the nonlinearity scale is comparable to the (Lagrangian) size
of halos, we can expect the leading higher-derivative term to be of similar order as the nonlinear
bias corrections. Consequently, terms involving even higher numbers of derivatives, for instance
$\nabla^4\delta$, must then be suppressed on the scales we are interested in. However, our approach in this paper is to precisely test this 
assumption, i.e. the relative importance of higher derivative bias compared to nonlinear
bias corrections with a wide range of biased tracers. 

Before we can proceed we need to address the renormalization of the bias parameters
\cite{McDonald:2006,Assassi:2014}. This issue arises when computing observable quantities (correlation
functions) based on the bias expansion above, as this leads to dependencies on the variance
$\sigma^2 = \left<\delta^2(\B{x})\right>$. The variance is sensitive to the highly nonlinear regime where
perturbation theory breaks down, but these sensitivities are not physical as they rely on the definition of the
bias parameters and, in fact, can be completely absorbed by appropriate redefinitions of the \emph{bare}
parameters appearing in Eq.~(\ref{eq:theory.dg_expansion}) \cite{McDonald:2006}. 

However, it was shown in
\cite{EggScoSmi1906} that a convenient way to circumvent this problem altogether is to recast the bias expansion in
terms of \emph{multi-point propagators}. In analogy with renormalized perturbation theory~\cite{CroSco0603a} and related approaches~\cite{BerCroSco0811}, the galaxy multi-point propagators $\Gamma_g^{(n)}$ are
defined as ensemble averaged derivatives of $\delta_g$ with respect to the linear matter density
$\delta_L$. This correspond to associating the (renormalized) bias parameters with sums of reducible diagrams with a fixed number of external lines (denoted by the number of derivatives). Written in Fourier space\footnote{For Fourier space integrals we use the convention
  $\delta(\B{x}) = \int_{\B{k}} \exp{(-i\B{k}\cdot\B{x})}\,\delta(\B{k})$ with the short-hand notation
  $\int_{\B{k}_1,\ldots,\B{k}_n} \equiv \int \text{d}^3k_1/(2\pi)^3 \cdots \text{d}^3k_n/(2\pi)^3$.} we thus have:
\begin{equation}
  \label{eq:theory.MPdef}
  \begin{split}
    \left<\frac{\partial\delta_g(\B{k})}{\partial\delta_L(\B{k}_1)\,\cdots\,\partial\delta_L(\B{k}_n)}\right>
    \equiv \;\, &(2\pi)^3\,\Gamma_g^{(n)}(\B{k}_1,\ldots,\B{k}_n) \\ &\times\,\delta_D(\B{k} - \B{k}_{1 \cdots n})\,,
  \end{split}
\end{equation}
where $\B{k}_{1 \cdots n} = \B{k}_1 + \ldots + \B{k}_n$. As the multi-point propagators are observables
themselves, they are automatically renormalized and act as the scale-dependent bias parameters in an expansion
of the form
\begin{align}
  \label{eq:theory.dg_MPexpansion}
  \delta_g = \Gamma_g^{(1)} \otimes {\cal H}_1 + \Gamma_g^{(2)} \otimes {\cal H}_2 + \ldots\,,
\end{align}
which is equivalent to Eq.~(\ref{eq:theory.dg_expansion}) \emph{after} the usual renormalization procedure. The
product $\otimes$ is given by
\begin{equation}
  \label{eq:theory.prod_def}
  \begin{split}
    \left[\Gamma_g^{(n)} \otimes {\cal H}_n\right](\B{k}) \equiv \;\, &(2\pi)^3 \int_{\B{k}_1,\ldots,\B{k}_n}
    \delta_D(\B{k}-\B{k}_{1 \cdots n}) \\ &\times\,\Gamma_g^{(n)}(\B{k}_1,\ldots,\B{k}_n)\,{\cal
      H}_n(\B{k}_1,\ldots,\B{k}_n) \,,
  \end{split}
\end{equation}
and the ${\cal H}_n$ are the Wiener-Hermite functionals \cite{Matsubara:1995,EggScoSmi1906}, whose first two
representatives are ${\cal H}_1 = \delta_L^*(\B{k})$ and
${\cal H}_2 = \delta_L^*(\B{k}_1)\,\delta_L^*(\B{k}_2) - \left<\delta_L(\B{k}_1)\,\delta_L(\B{k}_2)\right>$. The
scale dependence of the multi-point propagators can be predicted based on the various bias contributions and
their nonlinear evolution. Up to the order we are working at one can show that \cite{EggScoSmi1906}
\begin{align}
  \label{eq:theory.MP1}
  \Gamma_g^{(1)}(\B{k}) = \;\, &b_1\left[1 + 3 \int_{\B{q}}F_3(\B{k},\B{q},-\B{q})\,P_L(q)\right] - \beta_1\,k^2
                                 \nonumber \\ &+
  4\gamma_2\int_{\B{q}} K(\B{k}-\B{q},\B{q})\,G_2(\B{k},-\B{q})\,P_L(q) \nonumber \\ &+ 2\gamma_{21}\int_{\B{q}}
  K(\B{k}-\B{q},\B{q})\,K(\B{k},\B{q})\,P_L(q)\,,
\end{align}
at one loop, as well as
\begin{equation}
  \label{eq:theory.MP2}
  \Gamma_g^{(2)}(\B{k}_1,\B{k}_2) = 2b_1\,F_2(\B{k}_1,\B{k}_2) + b_2 + 2\gamma_2\,K(\B{k}_1,\B{k}_2)\,,
\end{equation}
to tree level, where $P_L$ is the linear matter power spectrum defined below, while $F_n$ and $G_n$ denote the standard
perturbation theory (SPT) kernels (see e.g.~\cite{Bernardeau:2002}), and
\mbox{$K(\B{k}_1,\B{k}_2) = (\B{k}_1 \cdot \B{k}_2/k_1\,k_2)^2-1$} is the Fourier transform of the kernel describing 
${\cal G}_2(\Phi_v)$. The bias parameters in Eqs.~(\ref{eq:theory.MP1}) and (\ref{eq:theory.MP2}) are now the
renormalized ones (indicated by the lack of an over-bar) and their relation to the bare parameters can be
computed if desired \cite{EggScoSmi1906}. We stress that the expressions for the multi-point propagators are free
from contributions involving $\sigma^2$. Propagators for the nonlinear matter field, $\Gamma_m^{(n)}$, can be
derived from the expressions above in the limit that $b_1 \to 1$ and all other bias parameters are set to zero, e.g. Eq.~(\ref{eq:theory.MP1}) gives the well known one-loop propagator $\Gamma_m^{(1)}=1+P_{13}/(2P_L)$, with $P_{13}$ the only one-loop power spectrum reducible diagram in SPT~\cite{Bernardeau:2002}.

\subsection{Power Spectra}
\label{sec:power-spectra}

We are interested in the auto and cross power spectra of galaxies and matter, which are defined as follows
\begin{align}
  \left<\delta_g(\B{k})\,\delta_g(\B{k}')\right> &\equiv (2\pi)^3\,P_{gg}(k)\,\delta_D(\B{k}+\B{k}')\,, \\ 
  \left<\delta_g(\B{k})\,\delta(\B{k}')\right> &\equiv (2\pi)^3\,P_{gm}(k)\,\delta_D(\B{k}+\B{k}')\,.
\end{align}
When $\delta$ and $\delta_g$ are expressed in terms of multi-point propagators, we can exploit the orthogonality
of the Wiener-Hermite functionals \cite{Matsubara:1995,EggScoSmi1906} to immediately obtain an expression for the
cross power spectrum at one-loop order:
\begin{equation}
  \label{eq:theory.Pgm_1loop_MP}
  \begin{split}
    P_{gm}(k) = \;\, &\Gamma_g^{(1)}(k)\,\Gamma_m^{(1)}(k)\,P_L(k) + \frac{1}{2}\int_{\B{q}}
    \Gamma_g^{(2)}(\B{k}-\B{q},\B{q}) \\ &\times\,\Gamma_m^{(2)}(\B{k}-\B{q},\B{q})\,P_L(|\B{k}-\B{q}|)\,P_L(q)\,,
  \end{split}
\end{equation}
keeping only terms that are at most quadratic in $P_L$. An analogous result holds for $P_{gg}(k)$ and upon
inserting Eqs.~(\ref{eq:theory.MP1}) and (\ref{eq:theory.MP2}) we get
\begin{widetext}
  \begin{align}
    P_{gm}(k) &= b_1\,P_{mm}(k) - \beta_1\,k^2\,P_L(k) + 2 P_L(k) \int_{\B{q}}
                \Big[2\gamma_2\,G_2(\B{k},-\B{q}) + \gamma_{21}\,K(\B{k},\B{q})\Big]
                \,K(\B{k}-\B{q},\B{q})\,P_L(|\B{k}-\B{q}|) \nonumber \\ & \hspace{1.5em}
                + \int_{\B{q}} \Big[b_2+2\gamma_2\,K(\B{k}-\B{q},\B{q})\Big]
                \,F_2(\B{k}-\B{q},\B{q})\,P_L(|\B{k}-\B{q}|)\,P_L(q)\,, \label{eq:theory.Pgm_1loop} \\
    P_{gg}(k) &= 2b_1\,P_{gm}(k) - b_1^2\,P_{mm}(k) + \frac{1}{2}\int_{\B{q}} \Big[b_2^2 +
                2b_2\,\gamma_2\,K(\B{k}-\B{q},\B{q}) 
                +4\gamma_2^2\,K(\B{k}-\B{q},\B{q})^2\Big]\,P_L(|\B{k}-\B{q}|)\,P_L(q)\,, \label{eq:theory.Pgg_1loop}
  \end{align}
\end{widetext}
where $P_{mm}(k)$ is the one-loop nonlinear matter power spectrum.

\subsection{Matter modeling}
\label{sec:matter}

The dominant contribution to Eqs.~(\ref{eq:theory.Pgm_1loop}) and (\ref{eq:theory.Pgg_1loop}) in the nonlinear
regime typically comes from the nonlinear matter power spectrum, i.e., the term that is multiplied by either
$b_1$ or $b_1^2$ in $P_{gm}$ and $P_{gg}$, respectively. Its correct modeling is therefore of particular
importance and it has been shown to great detail that the SPT matter power spectrum that derives from the
expressions above is subject to sizable inaccuracies, which spoil its agreement with measurements from
simulations. This can be attributed to two main reasons: first, SPT does not properly take into account the
effect from large-scale relative displacements, which give rise to a damping of the baryon acoustic
oscillation (BAO) signature. Second, a further simplification made in the context
of SPT is the assumption that the dark matter field behaves as a pressureless perfect fluid, which implies a
vanishing stress-tensor, that can be described perturbatively. While this is a good approximation on sufficiently large scales, it is no longer
applicable in the nonlinear regime and therefore needs to be accounted for by
appropriate correction terms.

\subsubsection{BAO damping from large-scale ``infrared" modes}
\label{sec:BAO}

It has long been  recognized that large-scale modes are responsible for the damping of the amplitude of BAO and can thus be treated perturbatively~\cite{CroSco0603b,EisSeoWhi0708,EisSeoSir0708,MatPie08,CroSco0801}. The relevant quantity for the power spectrum is the relative displacement field  two-point function at the BAO scale, which smears the BAO signal and receives relatively large corrections from large-scale modes. 

The large-scale displacement field gives the most important contribution to the decay of the matter propagator, and these contributions were first resummed in the context of renormalized perturbation theory (RPT)~\cite{CroSco0603b} (see also~\cite{MatPie0706,Mat0803,BerCroSco1206,TarBerNis1211}) and applied  to BAO in~\cite{CroSco0801}.  Another useful way to think about BAO damping at large scales is to decompose the linear power spectrum  into a smooth ($P_{\mathrm{nw}}$) and wiggly component\footnote{There exist different methods in
  the literature that aim to isolate the wiggly component from a given linear power spectrum. For our explicit
  computations in Section~\ref{sec:results_models} we chose to implement the Gaussian smoothing method discussed
  in Appendix A of \cite{VlaSelChu1603}.} 
($P_{\mathrm{w}}$), such that~\cite{SeoSieEis0810}
\begin{equation}
  P_L(k) = P_{\mathrm{nw}}(k) + P_{\mathrm{w}}(k)\,.
  \label{eq:PLdecomp}
\end{equation}
then the effect of  relative displacements is to ``dewiggle" the spectrum, which can be described to leading order in PT by applying the smearing operator in the Zel'dovich approximation $\exp(-k^2 \Sigma^2)$ to the wiggly part, 
\begin{equation}
  P_{L,\mathrm{dewiggled}}(k) = P_{\mathrm{nw}}(k) +  \text{e}^{-k^2\,\Sigma^2}\,P_{\mathrm{w}}(k)\,, \label{eq:dewiggled} 
 \end{equation}
where $\Sigma^2$ is the relative displacement field two-point function~\cite{EisSeoWhi0708} at the BAO scale. These two approaches (resummation and dewiggling) have been brought together and put on a much sounder footing more recently, as we now discuss. 

RPT and related methods, e.g. RegPT~\citep{BerCroSco1206,TarBerNis1211}, resum the propagator but keep the mode-coupling contributions to the power spectrum at fixed order, and this asymmetry leads to breaking of the Galilean invariance (GI)\footnote{More precisely, here GI should be thought as a more general invariance under uniform  velocity or displacement fields with arbitrary time-dependence~\cite{PelPie1302,KehRio1308}, i.e. including accelerations, which can be related to the equivalence principle~\cite{CreNorSim1312,CreGelSim1402}. } of the equal-time correlators~\cite{ScoFri9607}, which results in unphysical damping of the broadband power, particularly at large $k$. This can be avoided to a large extent, a posteriori, by requiring that the power spectrum be invariant under large-scale displacements, effectively constructing a resummation of the mode-coupling contributions consistent with that of the propagator. \correction{Starting from the RegPT form of the propagator to one-loop~\citep{BerCroSco1206},
\beq
G= \Big( 1 + {P_{13}^{\rm inv} \over 2P_L} \Big) \, {\rm e}^{-k^2 \sigma_v^2/2}
\label{GregPT}
\eeq
where $P_{13}^{\rm inv} \equiv P_{13}+k^2 \sigma_v^2 P_L$ is the invariant (under large-scale displacements)
component of $P_{13}$ and $\sigma_v^2$ is the linear velocity (or displacement) variance given by
$\sigma_v^2=\int_{\bf q} P_L(q)/3$, one can derive the transformation properties of the one-loop power spectrum $P_{mm} = G^2 P_L + P_{22}$~\cite{CroSco0603a}. Under a uniform random displacement with variance $\sigma_u^2$, it follows that $\sigma_v^2 \to \sigma_v^2 + \sigma_u^2$, $P_{13} \to P_{13} - k^2 \sigma_u^2P_L$ and $P_{22} \to P_{22} + k^2 \sigma_u^2P_L$. Requiring that the power spectrum be invariant, $\partial P_{mm}/\partial \sigma^2_u=0$,  gives an expression for $\sigma_u^2$ at each $k$, resulting in 
\beq
P_{mm,\,\mathrm{gRPT}}(k)= \Big( 1 + { P_{13}^{\rm inv} \over P_L} \Big) \, {\rm e}^{x} \Big[ (1-x)\, P_L + P_{22}^{\rm inv} \Big]
\label{Pgrpt1L}
\eeq
where  $P_{22}^{\rm inv} \equiv P_{22}-k^2 \sigma_v^2 P_L$. The invariance condition gives $x P_L = P_{1\ell} - \sqrt{ P_{1\ell}^2 - 2P_LP_{2\ell}}$, with $P_{n\ell}$ being the $n$-loop SPT power contribution. Since we are calculating a relative large-scale displacement, for simplicity we use the Zel'dovich approximation to compute $P_{n\ell}$ inside $x$, which avoids issues related to the UV sensitivity of the 2-loop propagator in the exact dynamics.}

\correction{All ingredients in Eq.~(\ref{Pgrpt1L}), $P_{13}^{\rm inv},x,P_{22}^{\rm inv}$ are invariant under large-scale displacements. 
However this procedure, known as gRPT,  cannot generate all invariants in the mode-coupling contributions as they are not required to cancel the non-invariants from the propagator contributions. We will implement Eq.~(\ref{Pgrpt1L}), already used in~\cite{SanScoCro1701,TroSanAsg2001}, as one of the ways to compute the nonlinear matter contributions in Section~\ref{sec:results_models}. See Fig~2 in~\cite{SanScoCro1701} for a comparison of the one-loop power spectrum against simulations at $z=0.57$, in the middle of the redshift range that concerns us here. For a similar approach to invariant resummations see~\cite{PelPie1609}. }

Generalizing such resummations to preserve GI throughout is technically cumbersome, but has been done
in~\cite{SenZal1502} by mixing Eulerian PT with the Lagrangian description to compute the large-scale relative
displacements effects on the density power spectrum BAO. This resummation procedure for the equal-time two-point
correlator reduces to RegPT (restricted to infrared modes as in~\cite{BervanVer1203}) when one of the fields is
at the initial conditions, i.e. when applied to the propagator, as expected. Taking advantage of the fact that
large-scale relative displacements smear sharp features such as the BAO but not the broadband shape,
\cite{BalMirSim1508} and \cite{BlaGarIva1607} show that one can formulate the resummation of infrared modes
systematically directly in Fourier space by using the decomposition in Eq.~(\ref{eq:PLdecomp}).  The full
expressions of the infrared resummed matter power spectrum at linear and to one-loop order are then given
by~\cite{BalMirSim1508}
\begin{align}
  P_{L,\mathrm{IR}}(k) &= P_{\mathrm{nw}}(k) +  \text{e}^{-k^2\,\Sigma^2}\,P_{\mathrm{w}}(k)\,, \label{eq:theory.PLIR} \\
  P_{mm,\mathrm{IR}}(k) &= P_{L,\mathrm{IR}}(k) + P_{\mathrm{nw},\mathrm{1-loop}}(k) \nonumber \\
                       &\hspace{1.2em}+ \text{e}^{-k^2\,\Sigma^2} \Big[k^2\,\Sigma^2\,P_{\mathrm{w}}(k) +
                         P_{\mathrm{w},\mathrm{1-loop}}\Big]\,, \label{eq:theory.PmmIR}
\end{align}
where $P_{\mathrm{nw},\mathrm{1-loop}}$ is the SPT one-loop correction but evaluated using the smooth linear
spectrum $P_{\mathrm{nw}}$ and $P_{\mathrm{w},\mathrm{1-loop}} = P_{\mathrm{SPT},\mathrm{1-loop}} -
P_{\mathrm{nw},\mathrm{1-loop}}$. The first term in the square bracket of Eq.~(\ref{eq:theory.PmmIR}) guarantees that the limit of small $\Sigma^2$ reduces to one-loop SPT, as it should. The damping kernel is given by the relative displacement two-point function computed in the Zel'dovich approximation at the BAO scale~\cite{EisSeoWhi0708}:
\begin{equation}
  \label{eq:theory.BAOdampfactor}
  \Sigma^2 = \int_0^{k_S} \hspace{-0.5em}{\text{d}^3q\over (2\pi)^3}\,{P_{\mathrm{nw}}(q)\over 3\,q^2} \left[1 -
    j_0\left(\frac{q}{k_{\mathrm{\tiny BAO}}}\right) + 2j_2\left(\frac{q}{k_{\mathrm{\tiny BAO}}}\right)\right]\,,
\end{equation}
where $j_n(x)$ are spherical Bessel functions of order $n$ and $k_{\mathrm{\tiny BAO}}$ corresponds to the BAO scale
in frequency space, i.e., $k_{\mathrm{\tiny BAO}} = \pi/\ell_{\mathrm{\tiny BAO}}$ with BAO scale taken as a fixed scale $\ell_{\mathrm{\tiny BAO}} =
110\,\Mpc$. For $q\ll k_{\mathrm{\tiny BAO}}$ the square brackets suppress the integrand by $(q/k_{\mathrm{\tiny BAO}})^2$, as expected since displacements longer than $\ell_{\mathrm{\tiny BAO}}$ do not contribute to the smear of the BAO, although that suppression does not affect  the value of $\Sigma^2$ in practice due to the shape of the CDM spectrum. For $q\gg k_{\mathrm{\tiny BAO}}$ the square brackets are dominated by one-point displacement (first term). The other scale that enters Eq.~(\ref{eq:theory.BAOdampfactor}) is $k_S$, which serves to separate
large-scale from small-scale modes and thus indicates the range of scales that are being resummed. Strictly speaking, to perform an infrared resummation one should sum over modes $q<k$, but in practice this cutoff is typically set to the fixed value $k_S = 0.2\,\iMpc$ e.g. \cite{BlaGarIva1607,IvaSimZal1909}, and we will follow this practice below in Section~\ref{sec:results_models}. Equations~(\ref{eq:theory.PLIR}-\ref{eq:theory.PmmIR}) then represent the infrared resummed one-loop power spectrum in SPT.

\subsubsection{Small-scale corrections: non-trivial stress tensor}
\label{sec:EFT}

Orbit crossing generates a non-zero stress tensor even if dark matter is perfectly cold to begin with, i.e. if it has an initial distribution function corresponding to a delta function in momentum space. The width of this distribution function is characterized by the stress tensor ${\bold \sigma}$, which gets generated by dynamics in multistreaming regions~\cite{PueSco0908}. Estimates from numerical simulations show that for CDM spectra this effect is small compared to loops but not altogether negligible on large scales~\cite{PueSco0908,PieManSav1108,NodPelPie1708,BueHah1907}. The leading order correction to the power spectrum is to add a $k^2 P(k)$ suppression~\cite{PueSco0908,CarHerSen1206,BauNicSen1207}, 
\begin{equation}
  \label{eq:theory.counterterm}
  P_{\sigma}(k) = -2 c_1^2\,k^2\,P_L(k)\,,
\end{equation}
with $c_1$ being a number that is linked to the time-dependence of the stress-tensor and occasionally referred
to as \emph{effective speed of sound} in the context of effective field theory (EFT,~\cite{CarHerSen1206,BauNicSen1207}). The time-dependence of the stress-tensor is difficult to calculate for generic 
$\Lambda$CDM universes and so $c_1$ has to be treated as an additional free parameter that is determined from
the data itself (see e.g.~\cite{CarForGre1310,BalMerZal1512}). Moreover, Eq.~(\ref{eq:theory.counterterm}) satisfies a second purpose: it allows to absorb the
leading sensitivities to the highly nonlinear regime, which can arise as the loop integrals are nominally
performed over the full range of scales, including those where we do not expect perturbation theory to hold. For
typical linear $\Lambda$CDM power spectra these integrals converge rapidly, which implies that any such
sensitivities must be dominated by the lowest order term in the high-$q$ expansion\footnote{By $q$ we mean the
  loop momentum that is integrated over, see e.g. Eqs.~(\ref{eq:theory.Pgm_1loop}) and
  (\ref{eq:theory.Pgg_1loop}).} of a given loop integral
\cite{Bernardeau:2012,PajZal1308,BlaGarKon1309,BerTarNis1401,BalMerZal1512}, and which scales as
\mbox{$\sim k^2\,P_L(k)$}.

Combining the IR resummed one-loop SPT spectrum with these small-scale corrections, the total  matter power spectrum can thus be written as
\begin{equation}
  \label{eq:theory.PmmEFT}
  P_{mm,\mathrm{EFT}}(k) = P_{mm,\mathrm{IR}}(k) - 2c_1^2\,k^2\,P_{L,\mathrm{IR}}(k)\,,
\end{equation}
where we denote this  as the EFT one-loop matter spectrum, and we have replaced the linear power spectrum from Eq.~(\ref{eq:theory.counterterm}) with its infrared resummed equivalent\footnote{In principle, the IR resummation should also be applied to the galaxy bias loop
  corrections. As we leave cosmological parameters (that have the strongest impact on the BAO feature) fixed in
  the model fits later on in this work, it should make negligible difference if this correction is being
  ignored. We have explicitly verified that this is indeed the case.}. Similarly, one should add $P_{\sigma}$ to the gRPT predictions discussed above, but as mentioned the broadband gRPT power is a bit lower at high-$k$ compared to SPT, which means that $c_1$ would be even a smaller correction in this case,  therefore we shall set $c_1=0$ for simplicity.  This is also reasonable since we are ultimately interested here in computing the power spectrum of biased tracers, and bias loops also change broadband power; in particular, the $\gamma_{21}$ bias contribution is \emph{exactly degenerate} with $P_\sigma$ in the low-$k$ limit~\cite{SanScoCro1701}. We discuss this issue further in Section~\ref{sec:results_models}.

As a third approach beyond gRPT and EFT, we  consider the combination of a simulation calibrated power
spectrum at a fixed cosmology with a perturbative model based on RegPT for the so-called \emph{response function}. The response
function $K(k,q)$ was introduced in \cite{BerTarNis1401,NisBerTar1611} and quantifies the variation of the nonlinear matter
power spectrum at scale $k$ induced by a change of the linear power at scale $q$. More precisely and in
similarity to the multipoint propagators, it is defined as a functional derivative of $P_{mm}(k)$ with respect
to $P_L(q)$,
\begin{equation}
  \label{eq:theory.response_function}
  K(k,q) \equiv q\,\frac{\partial P_{mm}(k)}{\partial P_L(q)}\,.
\end{equation}
A model for the response function using RegPT at two-loop order was presented in \cite{NisBerTar1712} and compared
to measurements from simulations. This study revealed very good agreement for $k\simeq q$, but not as good otherwise. The deviations $q\ll k$ are due to the breaking of GI in the RegPT resummation as discussed above, whereas for $q \gg k$ the response function had too strong sensitivity to small scale modes, a sign of a non-trivial stress tensor that is neglected in RegPT. Based on these considerations, the authors of~\cite{NisBerTar1712} provided a phenomenologically motivated modification to their model (without the need for additional free parameters) that was subsequently shown to reproduce the simulation measurements over a wide range of scales for $k$ and $q$. 

The response function approach becomes particularly powerful once the matter power spectrum for some fiducial
set of cosmological parameters $\B{\theta}_{\mathrm{fid}}$ is known with high precision since the response
function allows us to translate this to another cosmology $\B{\theta}$ as follows:
\begin{equation}
  \label{eq:theory.respresso}
  \begin{split}
    P_{mm}(k | \B{\theta}) =\,&P_{mm}(k | \B{\theta}_{\mathrm{fid}}) + \int \mathrm{d} \ln{q}\,K(k,q) \\
    &\times \left[P_L(q | \B{\theta}) - P_L(q | \B{\theta}_{\mathrm{fid}})\right]\,.
  \end{split}
\end{equation}
This expression is only valid if the difference between the linear power spectra of the fiducial and target
cosmologies is small, but as shown in \cite{NisBerTar1712} this limitation can be overcome by performing a
multi-step reconstruction that takes into account the cosmology dependence of the response function. This
procedure has been implemented in the \texttt{RESPRESSO} package \cite{NisBerTar1712}, which is making use of a
fiducial matter power spectrum at the Planck 2015 cosmology \cite{Planck1609} obtained from a set of
high-resolution simulations with suppressed variance \cite{AngPon1610}. \correction{Currently \texttt{RESPRESSO}
  is limited to the prediction of the nonlinear matter power spectrum, while an application to redshift-space
  measurements would require an extension to the matter-velocity and velocity-velocity power spectra. Their
  response functions can be defined in analogy with Eq.~(\ref{eq:theory.response_function}), but a detailed
  computation along the lines of \cite{NisBerTar1712} remains to be done. We also note that the response
  function approach is not limited to $\Lambda$CDM cosmologies and corrections, e.g. due to massive neutrinos,
  can be accounted for in the model of the response function in the same way they would be included in the gRPT
  or EFT models discussed above (see e.g. \cite{SaiTakTar0805,DupBer1401,BlaGarKon1411}).}

Summarizing, in Section~\ref{sec:results_models} we shall compute the nonlinear matter spectrum using gRPT, EFT and \texttt{RESPRESSO}  and check whether our results on one-loop bias are sensitive to this choice, in particular when compared to using the measured nonlinear matter spectrum in the simulations as done in Section~\ref{sec:results_true_matter} (our baseline, main results). Note that these three models are on a somewhat unequal footing, in the sense that gRPT and \texttt{RESPRESSO} have the same number of free parameters, but the latter has in it information from simulations already; on the other hand, EFT has one extra free parameter compared to the other two that should help fit the data, at the expense of perhaps a lower figure of merit. That provides a useful range of strategies that can be applied to data.

\subsubsection{Degeneracy between stress-tensor and \\ higher-derivative effects} 
\label{sec:stress-hd-degeneracy}

The counter-term due to a non-vanishing stress-tensor is clearly degenerate with the higher-derivative bias
contribution that was discussed in Section~\ref{sec:galaxy-bias}. However, since at one-loop order the
counter-term enters only through $P_{mm}$, the two effects leave different imprints on the auto and cross galaxy
power spectrum:
\begin{align}
  P_{gm}(k) &\supset -(2b_1\,c_1 + \beta_1)\,k^2\,P_L(k) \equiv -\beta_P^{\times}\,k^2\,P_L(k)\,, \label{eq:theory.Pgmk2} \\
  P_{gg}(k) &\supset -2b_1\,(b_1\,c_1 + \beta_1)\,k^2\,P_L(k) \equiv -\beta_P\,k^2\,P_L(k)\,, \label{eq:theory.Pggk2}
\end{align}
which would in principle allow us to break their degeneracy and constrain $c_1$ and $\beta_1$ simultaneously in
a joint analysis, for instance, in combinations of galaxy clustering and weak lensing. This ceases to be true if
there are significant sensitivities to the nonlinear regime stemming from the bias loop integrals, which would
also be absorbed by the $k^2\,P_L(k)$ dependent terms. As they are not guaranteed to be identical for $P_{gm}$
and $P_{gg}$, this would lead to different values of $c_1$ in Eqs.~(\ref{eq:theory.Pgmk2}) and
(\ref{eq:theory.Pggk2}). For that reason we choose the more conservative approach of collectively describing
these effects by two independent parameters, $\beta_P^{\times}$ and $\beta_P$, when performing model fits. If,
on the other hand, only higher-derivative bias is relevant we note that in that case we would have
$\beta_P = 2b_1\,\beta_P^{\times}$. This is what happens, by definition, when we use the nonlinear matter spectrum measured from simulations to test one-loop galaxy bias in Section~\ref{sec:results_true_matter} rather than the models described above, and thus it is a useful consistency check.

\subsection{Stochasticity}
\label{sec:stochasticity}

We now turn to the final ingredient in modeling the clustering of biased tracers, namely, how their formation
history is impacted by very short-wavelength fluctuations. While the description of these modes is beyond the
conventional reach of perturbation theory, an important characteristic is that they are mostly uncorrelated with
the long-wavelength perturbations. That means on large scales they can only contribute as a stochastic
field\footnote{Gravitational evolution leads to couplings between long and short wavelength perturbations, which
  requires the existence of additional composite operators between stochastic fields and terms of the general
  bias expansion, such as $\delta$ etc. For the power spectrum at one-loop order these extra fields are of
  subleading importance, but they give a relevant contribution to the bispectrum \cite{Desjacques:2018}.}
$\varepsilon_g$, which does not correlate over long distances and is thus described by highly localized
$N$-point functions in configuration space. In Fourier space this allows for the following expansion of the
stochasticity power spectrum $C_{gg}(k)$ \cite{Desjacques:2018}:
\begin{equation}
  \label{eq:theory.Pstoch}
  \begin{split}
    \left<\varepsilon_g(\B{k})\,\varepsilon_g(\B{k}')\right> &=
    (2\pi)^3\,C_{gg}(k)\,\delta_D(\B{k}+\B{k}') \\ &= (2\pi)^3 \left[N_0 + N_2\,k^2 +
      \ldots\right]\,\delta_D(\B{k}+\B{k}')\,,
  \end{split}
\end{equation}
with the constant term, $N_0$, representing deviations from purely Poissonian shot noise on large scales. The
scale-dependent piece is meant to account for the leading short-range nonlocality of the stochastic field in
analogy with the higher-derivative effects introduced above. Consequently, the parameter $N_2$ is also
associated to some intrinsic length scale tied to the Lagrangian radius of the dark matter
halos\footnote{Throughout this paper we will refer to this interchangeably as scale-dependent stochasticity or
  scale-dependent noise.}.

These shot noise corrections are qualitatively expected from the halo exclusion effect that arises from the
condition that dark matter halos cannot overlap, which implies that their correlation function must approach
$-1$ on scales below their radii~\cite{MoWhi9609,SheLem99,SmiScoShe0703,BalSelSmi1310}. 
As has been shown in \cite{Sch1603}, for the power spectrum this leads to a
modification of Poisson shot noise on large scales that transitions to the familiar value $1/\bar{n}$ in the
limit $k \to \infty$ ($\bar{n}$ being the average tracer number density).  Depending on the tracer
of the matter field, $N_0$ can either be negative or positive, indicating large-scale sub- or super-Poissonian shot noise,
respectively. The former occurs usually for galaxy populations with low satellite fractions that trace the
centers of massive halos, whereas super-Poisson values have been observed in simulations for low-mass subhalos
or galaxies that frequently appear as satellites \cite{CasMoShe0207,BalSelSmi1310}.

While the first scale-dependent term in Eq.~(\ref{eq:theory.Pstoch}) is able to capture an emerging change in
the shot noise value with increasing $k$, we stress that it needs to be strictly interpreted as a low-$k$
expansion because the truncation at finite order introduces various evident shortcomings: 1) in the limit
$k \to \infty$ Eq.~(\ref{eq:theory.Pstoch}) does not approach Poisson noise as expected, and 2) it does not
converge upon Fourier transformation to configuration space, so that the description of analogous effects in the
correlation function cannot be described in this way. However, as long as one is interested in large scales, 
the nontrivial noise properties are all constrained to small scales in the correlation function, and therefore they can  be ignored.

The stochastic terms are also required from a different point of view. Similar to the loop integral's
sensitivity to the non-perturbative regime for growing values of $k$, they can also leave an impact on very
large scales. This becomes clear when taking the large-scale limit of Eq.~(\ref{eq:theory.Pgg_1loop}), which
gives
\begin{equation}
  \label{eq:theory.Pgg_limit}
  \lim_{k \to 0}\,P_{gg}(k) = \frac{b_2^2}{2} \int_{\B{q}} P_L(q)^2\,,
\end{equation}
but this value is highly dependent on the order at which we truncate the bias expansion as terms at two-loop
order and beyond will add to this limit \cite{EggScoSmi1906}. For that reason, we deal with such terms by
subtracting Eq.~(\ref{eq:theory.Pgg_limit}) from Eq.~(\ref{eq:theory.Pgg_1loop}) and absorb their contribution
into the free parameter $N_0$ \cite{McDonald:2006,McDonald:2009}. Subleading corrections to
Eq.~(\ref{eq:theory.Pgg_limit}), if relevant, scale as $\sim k^2$ and can similarly be absorbed by $N_2$.

Finally, it is also possible to have stochasticity in the cross power spectrum. While the field $\varepsilon_g$
cannot contribute since $\left<\varepsilon_g\,\delta\right> = 0$, it is expected that the matter field itself is
described by a deterministic part, captured by perturbation theory, and a stochastic part $\varepsilon_m$, which
is due to small-scale modes leaving an imprint on large scales. Such a field, on the other hand, would correlate
with $\varepsilon_g$ and on the grounds of mass and momentum conservation of the matter field the resulting
cross power spectrum
$\left<\varepsilon_g(\B{k})\,\varepsilon_m(\B{k}')\right> = (2\pi)^3 C_{gm}(k) \,
\delta_D(\B{k}+\B{k}')$ is suppressed by a factor of $k^2$
\cite{Pee80,SmiPeaJen0306,AboMirPaj1605,Desjacques:2018}. Therefore,
\begin{equation}
  \label{eq:theory.Pxstoch}
  C_{gm}(k) = N_2^{\times}\,k^2 + \ldots\,,
\end{equation}
which shows that stochasticity in the cross power spectrum is already a higher-order effect.

\subsection{Co-evolution relations}
\label{sec:co-evolution}

\begin{figure}
  \centering
  \includegraphics[width=\columnwidth]{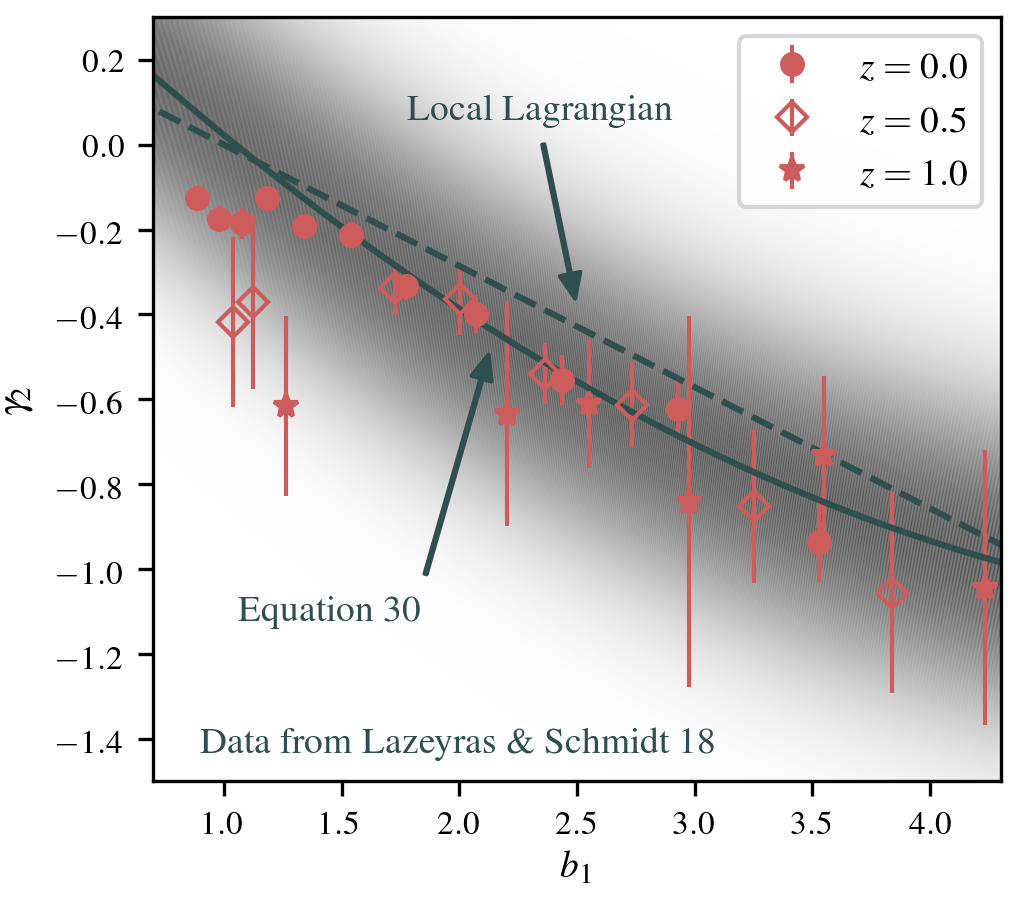}
  \caption{Comparison of the local Lagrangian relation for $\gamma_2$ and Eq.~(\ref{eq:theory.g2ECS}) to direct
    measurements at various redshifts from \cite{LazSch1712}. The shaded region denotes a Gaussian error band
    around Eq.~(\ref{eq:theory.g2ECS}) with width $0.25$, see text for details.}
  \label{fig:g2ECS}
\end{figure}

Nonlinear evolution contributes to the biasing of any class of tracers. It was shown in \cite{Fry9604,Chan:2012} that even when starting from a purely local relationship between the galaxy and matter density contrasts at the time of
formation (i.e., only the local bias parameters $b_n$ are unequal to zero), this is no longer the case at any
later point in time. Assuming conserved evolution (hereafter co-evolution) of the galaxies after formation 
one can derive the following co-evolution relations \cite{Chan:2012,Baldauf:2012,EggScoSmi1906}
\begin{align}
  \gamma_2 &= -\frac{2}{7}(b_1 - 1) + \gamma_{2,{\cal L}}\,, \label{eq:theory.g2evo} \\
  \gamma_{21} &= \frac{2}{21}(b_1 - 1) + \frac{6}{7}\gamma_2 + \gamma_{21,{\cal L}}\,, \label{eq:theory.g21evo}
\end{align}
where quantities with subscript ${\cal L}$ denote the \emph{Lagrangian} bias parameters, i.e. their values at
past infinity. The additional terms involved in Eqs.~(\ref{eq:theory.g2evo}) and (\ref{eq:theory.g21evo}) are
those induced by gravitational evolution and can be entirely expressed in terms of lower-order bias
parameters. We see that having zero nonlocal bias parameters at late times requires highly fine-tuned
Lagrangian parameters and various analyses that measured the tidal bias from simulated halo catalogs using
various techniques \cite{Chan:2012,Baldauf:2012,SheChaSco1304,SaiBalVla1405,LazSch1712,AbiBal1807} have conclusively ruled out
this scenario. On the other hand, the local Lagrangian approximation (LL), $\gamma_{2,{\cal L}} = 0$, provides a
much more accurate description of these measurements, even though the most recent results from
\cite{LazSch1712,AbiBal1807} indicate slightly lower values. This is shown in Fig.~\ref{fig:g2ECS}, where we
plot their data obtained from various halo samples at different redshifts against the LL prediction (dashed
line).

An alternative estimation (strictly speaking, not a co-evolution relation) of the tidal bias parameter in the
context of the excursion set approach was discussed in \cite{SheChaSco1304}. Using  a random walk with correlated steps to determine the probability for crossing the collapse barrier (see also~\cite{CasParHah1611,CasParShe1707}), they make a prediction for $\gamma_2$, which can be 
represented by the following quadratic fit:
\begin{equation}
  \label{eq:theory.g2ECS}
  \gamma_{2,\mathrm{ex}}(b_1) = 0.524 - 0.547\,b_1 + 0.046\,b_1^2\,.
\end{equation}
This fit is shown as the solid line in Fig.~\ref{fig:g2ECS} and provides a slightly better description of the
measurements than the LL assumption for $b_1 \gtrsim 1.3$. Hereafter we shall refer to
Eq.~(\ref{eq:theory.g2ECS}) as the ``excursion set relation" as a shorthand, but note this does not represent a
first-principle calculation. The shaded region around Eq.~(\ref{eq:theory.g2ECS}) in Fig.~\ref{fig:g2ECS}
denotes a Gaussian distribution with uncertainty of 0.25 around $\gamma_{2,\mathrm{ex}}$, which corresponds to
our best constraints in fits to be discussed below when $\gamma_2$ is set to be free. Given these uncertainties
the distinction between Eq.~(\ref{eq:theory.g2ECS}) and the LL relation is likely irrelevant for the purposes of
this work, but should be reconsidered when $\gamma_2$ is significantly better constrained, as is the case in
joint fits with the bispectrum \cite{Eggemeier}.

Relations such as the ones above can be useful because they allow us to reduce the number of free model
parameters that have to be marginalized over, which in turn restores constraining power on other, potentially
more interesting parameters. In general, fixing bias parameters via the co-evolution relations seems preferable
over ignoring them altogether, but one should proceed with caution, as this can still be a potential source of
bias in the analysis. In Section~\ref{sec:results_true_matter} we try to determine in detail for what range of
scales this is the case.

\section{Data and methodology}
\label{sec:data}

\begin{table*}
  \centering
  \caption{Labels and defining properties of the galaxy and halo samples used in our analysis. The last two
    columns denote the measurement of the linear bias and the large-scale deviation from Poisson shot noise (in
    units of $\bar{n}$ of the respective sample), as described in Section~\ref{sec:measurements_bias_noise}.}
  \begin{ruledtabular}
    \begin{tabular}{ccccccc}
      Identifier & Simulation & $z$ & $M_{\mathrm{halo}}$ [$10^{13}\,M_{\odot}$] & $\bar{n}$ [$\left(h/\mathrm{Mpc}\right)^3$] & $b_1$ 
      & $N_0$ \Tstrut\Bstrut\\ \hline
      MGS & \textsc{LasDamas} Carmen & $0.132$ & --- & $1.1 \times 10^{-3}$ & $1.414 \pm 0.003$ & $-0.16 \pm
                                                                                                  0.07$ \Tstrut\\
      LOWZ & \textsc{LasDamas} Oriana & $0.342$ & --- & $9.4 \times 10^{-5}$ & $2.235 \pm 0.012$ & $-0.176 \pm 
                                                                                                   0.018$ \\ 
      CMASS & \textsc{Minerva} & $0.57$ & --- & $4.0 \times 10^{-4}$ & $2.022 \pm 0.003$ & $-0.29 \pm 0.02$
                                                                                           \Bstrut\\ 
      \hline
      HALO1 & \textsc{LasDamas} Oriana & $0.0$ & $1$ - $10$ & $3.4 \times 10^{-4}$ & $1.442 \pm 0.004$ & $0.18
                                                                                          \pm
                                                                                          0.04$
                                                                                          \Tstrut\\
      HALO2 & \textsc{LasDamas} Oriana & $0.0$ & $> 10$ & $1.8 \times 10^{-5}$ & $2.92 \pm 0.02$ & $-0.35
                                                                                    \pm
                                                                                    0.02$
      \\
      HALO3 & \textsc{LasDamas} Oriana & $0.974$ & $1.3$ - $2$ & $5.1 \times 10^{-5}$ & $2.690 \pm 0.005$ & $-0.044
                                                                                             \pm
                                                                                             0.004$
      \\
      HALO4 & \textsc{LasDamas} Oriana & $0.974$ & $> 2$ & $5.0 \times 10^{-5}$ & $3.568 \pm 0.005$ & $-0.15
                                                                                       \pm
                                                                                       0.02$      \Bstrut \\
    \end{tabular}
  \end{ruledtabular}
  \label{tab:tracers}
\end{table*}

\begin{table}
  \centering
  \caption{Cosmological parameters of the dark matter simulations used to create our galaxy and halo
    catalogs.}
  \begin{ruledtabular}
    \begin{tabular}{ccccccc}
      Simulation & $\Omega_m$ & $\Omega_{\Lambda}$ & $\Omega_b$ & $h$ & $n_s$ & $\sigma_8$ \Tstrut\Bstrut\\
      \hline
      \textsc{LasDamas} & 0.25 & 0.75 & 0.04 & 0.7 & 1.0 & 0.8 \Tstrut \\
      \textsc{Minerva} & 0.285 & 0.715 & 0.046 & 0.695 & 0.9632 & 0.828 \Bstrut\\
    \end{tabular}
  \end{ruledtabular}
  \label{tab:cosmology}
\end{table}

\subsection{Galaxy and halo catalogs}
\label{sec:catalogs}

In order to test the performance of the previously described model for a diverse range of bias properties, we
use various populations of tracers at different redshifts, whose details are summarized in
Table~\ref{tab:tracers}.

We consider three galaxy catalogs at redshifts $z = 0.132$, $0.342$ and $0.57$ that were generated by
assigning galaxies to selected halos and subhalos of dark matter only N-body simulations according to a halo
occupation distribution (HOD). The catalogs for the lower two redshifts are based on the Carmen and Oriana boxes
of the \textsc{LasDamas} simulation suite~\cite{McBBerSco0901,SinBerMcB1807}, whose cosmology is defined by the parameters given in
Table~\ref{tab:cosmology}. The two boxes have a volume of $(1000\,\mathrm{Mpc}/h)^3$ and
$(2400\,\mathrm{Mpc}/h)^3$, respectively, mass resolution of $4.9 \times 10^{10}\,M_{\odot}/h$ and
$4.6 \times 10^{11}\,M_{\odot}/h$, and in total there are 40 realizations each that were set up with independent
initial conditions using CMBFAST~\cite{SelZal96} and second-order Lagrangian perturbation theory~\cite{CroPueSco0611}. The parameters of the HOD
model have been tuned for these two cases to match the number densities and projected two-point clustering of
the SDSS Main Galaxy Sample (MGS) with $M_r<-21$ and the BOSS LOWZ sample, see~\cite{SinBerMcB1807} for more details. The high-redshift galaxy catalog derives from the
\textsc{Minerva} simulations \cite{GriSanSal1604}, a set of 100 boxes of volume $(1500\,\mathrm{Mpc}/h)^3$  (see
Table~\ref{tab:cosmology} for cosmological parameters) with an HOD model that reproduces the properties of the
BOSS CMASS galaxies. Even though the volumes of the individual boxes are different from those of the actually
observed samples and our catalogs do not include the survey geometry or any systematic effects, for convenience
we will still refer to them as MGS, LOWZ and CMASS.

In addition, we identified two low- and two high-mass halo samples at redshifts $z = 0$ and $0.974$ from the
Oriana boxes. The low-mass bin at $z = 0$ contains halos of masses $1$ - $10\times 10^{13}\,M_{\odot}$, while
all of the more massive halos are grouped together in the high-mass bin. For the high redshift slice we
split halos according to $M_{\mathrm{halo}} \in [1.3,2]\times 10^{13}\,M_{\odot}$ and $M_{\mathrm{halo}} >
2\times 10^{13}\,M_{\odot}$. In the following we label these four samples as HALO1 to HALO4. All halos,
including the ones in the \textsc{Minerva} simulations on which the HOD's are based, were identified using the
friends-of-friends algorithm~\cite{DavEfsFre8505} with linking length equal to 0.2 times the mean interparticle
separation. The halos in the  \textsc{Minerva} simulations are then further processed through the subfind
algorithm~\cite{SprWanVog0812}.

\correction{Our galaxy samples and the applied mass cuts for the halo catalogs (which are limited by the
  resolution of the simulations) do not cover halo mass ranges below $\sim\,10^{12}\,M_{\odot}$. Even though the
  perturbative bias expansion should be universal, we thus caution that some of our results obtained in
  Section~\ref{sec:results_true_matter} (particularly the validity of co-evolution relations and the importance
  of higher-derivative and scale-dependent noise terms) do not necessarily generalize to the main targets of the
  DESI and Euclid missions. However, DESI will also observe large quantities of MGS- and BOSS-like galaxies (see
  \cite{DESI2}), as well as quasars, for which our study is of immediate relevance.}

\subsection{Measurements of power spectra and their covariances}
\label{sec:measurements_P_cov}

We measure the auto power spectrum, as well as the cross power spectrum with the nonlinear matter field, for
each of the galaxy and halo catalogs in Table~\ref{tab:tracers} using the estimator described in
\cite{SefCroSco1512}. For the galaxy samples these measurements are carried out in bins of $\Delta k = k_f$ over
the range $k_{\mathrm{min}} = \Delta k$ to $k_{\mathrm{max}} = 0.35\,\iMpc$, whereas we use $\Delta k = 2k_f$ and
an according range of scales for all four halo samples. Here, $k_f = 2\pi/L_{\mathrm{box}}$ denotes the
fundamental frequency of the simulation box, which is given by $6.3 \times 10^{-3}\,\iMpc$,
$2.6 \times 10^{-3}\,\iMpc$ and $4.2 \times 10^{-3}\,\iMpc$ for Carmen, Oriana and Minerva,
respectively. Finally, we correct the galaxy auto power spectra by subtracting the Poisson shot noise
contribution $1/\bar{n}$.

From the measurements over the $N_R$ independent realizations we further estimate the covariance matrices for
$P_{gg}$, $P_{gm}$ and their correlation as follows:
\begin{equation}
  C_{X \times Y,ij} = \frac{1}{N_R}\sum_{n=1}^{N_R} \left(X_i^{(n)}-\overline{X}_i\right)\,\left(Y_j^{(n)}-\overline{Y}_j\right)\,,
\end{equation}
where $X_i, Y_i = P_{gg}(k_i)$ or $P_{gm}(k_i)$ and over-bars denote their averages. As the number of
realizations is generally small compared to the total number of measurements, the estimated covariance matrices
are noisy or not even invertible in case their dimensionality exceeds $N_R$. However, with the exception of MGS,
all samples have such low number densities that the covariance matrices receive strong contributions from shot
noise, which boosts the diagonal elements and thus relatively decreases the correlation in the off-diagonal
terms. For that reason we only consider the diagonal part
of the covariance matrix when fitting $P_{gg}$ or $P_{gm}$ individually, but we retain the correlation between
the two power spectra when performing a joint analysis. To reduce noise we compare the measured uncertainties to the Gaussian prediction and retain the maximum between the two, which provides a conservative estimate. The resulting diagonal or block-diagonal matrices can be
inverted analytically.

\begin{figure}
  \centering
  \includegraphics{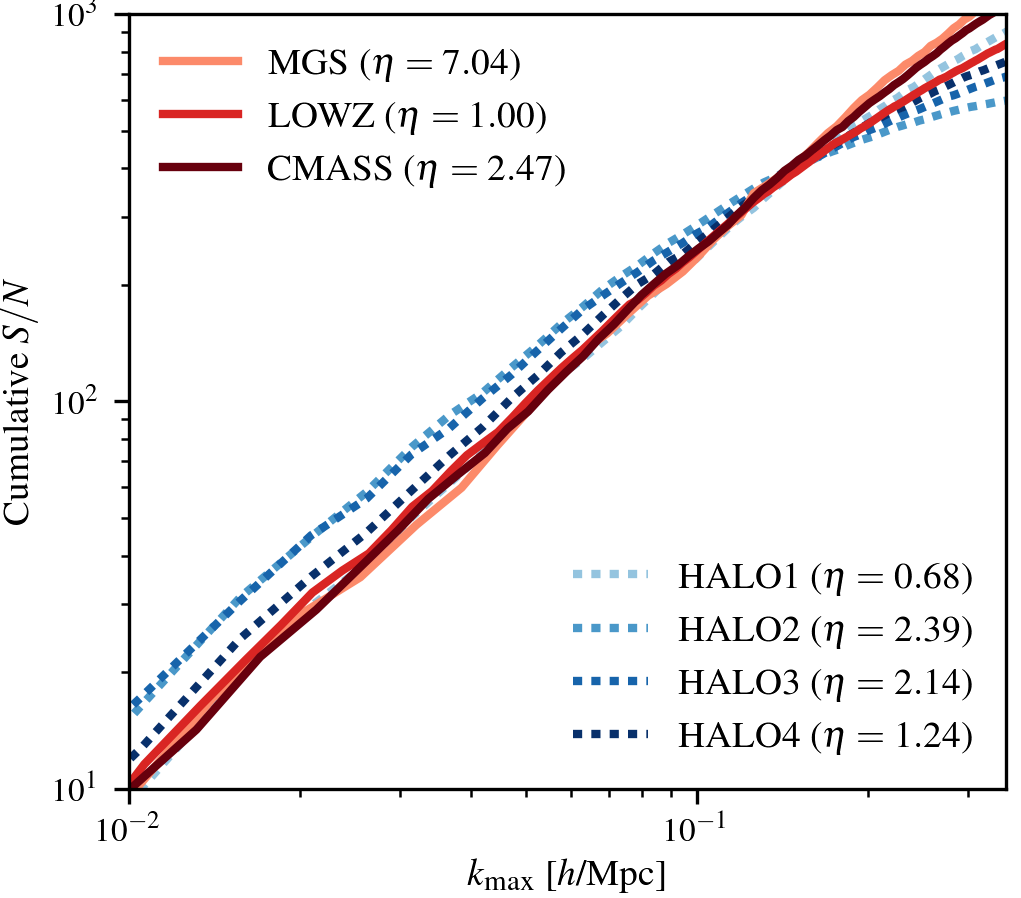}
  \caption{Cumulative signal-to-noise of galaxy and halo samples (solid and dotted lines, respectively) after
    rescaling the covariance matrices by the volume factor $\eta$ (see Eq.~\ref{eq:data.eta}).}
  \label{fig:cumulSN}
\end{figure}

As the various samples differ in terms of volume, number density and amplitude of the power spectrum, their
statistical power to constrain model parameters can also vary significantly. To reduce the effect of some of
these dependencies, we apply a volume scaling factor $\eta$ chosen such that all samples have identical
\emph{effective} volume \cite{FelKaiPea9405,Teg9711}, which is given by
\begin{equation}
  V_{\mathrm{eff}}(k) = \left[\frac{\bar{n}\,P_{gg}(k)}{1 + \bar{n}\,P_{gg}(k)}\right]^2\,V\,,
\end{equation}
when assuming a constant number density $\bar{n}$. We take the effective volume of the LOWZ galaxy sample
evaluated at $k = 0.1\,\iMpc$ as our reference, which yields
$V_{\mathrm{eff,LOWZ}} \approx 6\,(\mathrm{Gpc}/h)^3$. The scaling factor for all other samples is thus
determined by $\eta = V_{\mathrm{eff,LOWZ}}/V_{\mathrm{eff,X}}$ and as the covariance matrices at leading order
scale inversely with volume, we modify our estimates according to
\begin{equation}
  \label{eq:data.eta}
  C_{X,ij} \to C_{X,ij}/\eta\,.
\end{equation}
In Fig.~\ref{fig:cumulSN} we show the cumulative signal-to-noise of the various samples as a function of maximum
wavenumber $k_{\mathrm{max}}$ using these rescaled covariances (see Figure legend for the value of $\eta$ in each
case). While the differing levels of shot noise lead to a more or less strong suppression for high
$k_{\mathrm{max}}$, we note that the signal-to-noise is generally in good agreement over a large range of scales
relevant to our analysis.

\subsection{Measurements of the linear bias parameter and large-scale shot noise}
\label{sec:measurements_bias_noise}

The simulations allow us to make precise measurements of the linear bias and the non-Poissonian
correction to large-scale shot noise. As discussed in Section~\ref{sec:stochasticity} $P_{gm}$ is free of shot
noise in the limit $k \to 0$, so that taking the ratio with the matter auto power spectrum $P_{mm}$ (also
measured directly from the simulations in the same $k$-bins) recovers $b_1$:
\begin{equation}
  \label{eq:data.b1true}
  b_1 = \lim_{k \to 0} \frac{P_{gm}}{P_{mm}}\,.
\end{equation}
In practice we compute this ratio for each realisation before taking the average in order to cancel cosmic
variance and afterwards fit a constant to the first few bins. Specifically, we use the cutoff scales
$k_{\mathrm{max}} = 0.025\,\iMpc$, $0.028\,\iMpc$ and $0.021\,\iMpc$ for Carmen, Oriana and Minerva,
respectively. The resulting values of $b_1$ are given in Table~\ref{tab:tracers} and will be considered as the
\emph{ground truth}, when comparing to the output of the model fits below.

\begin{figure}
  \centering
  \includegraphics{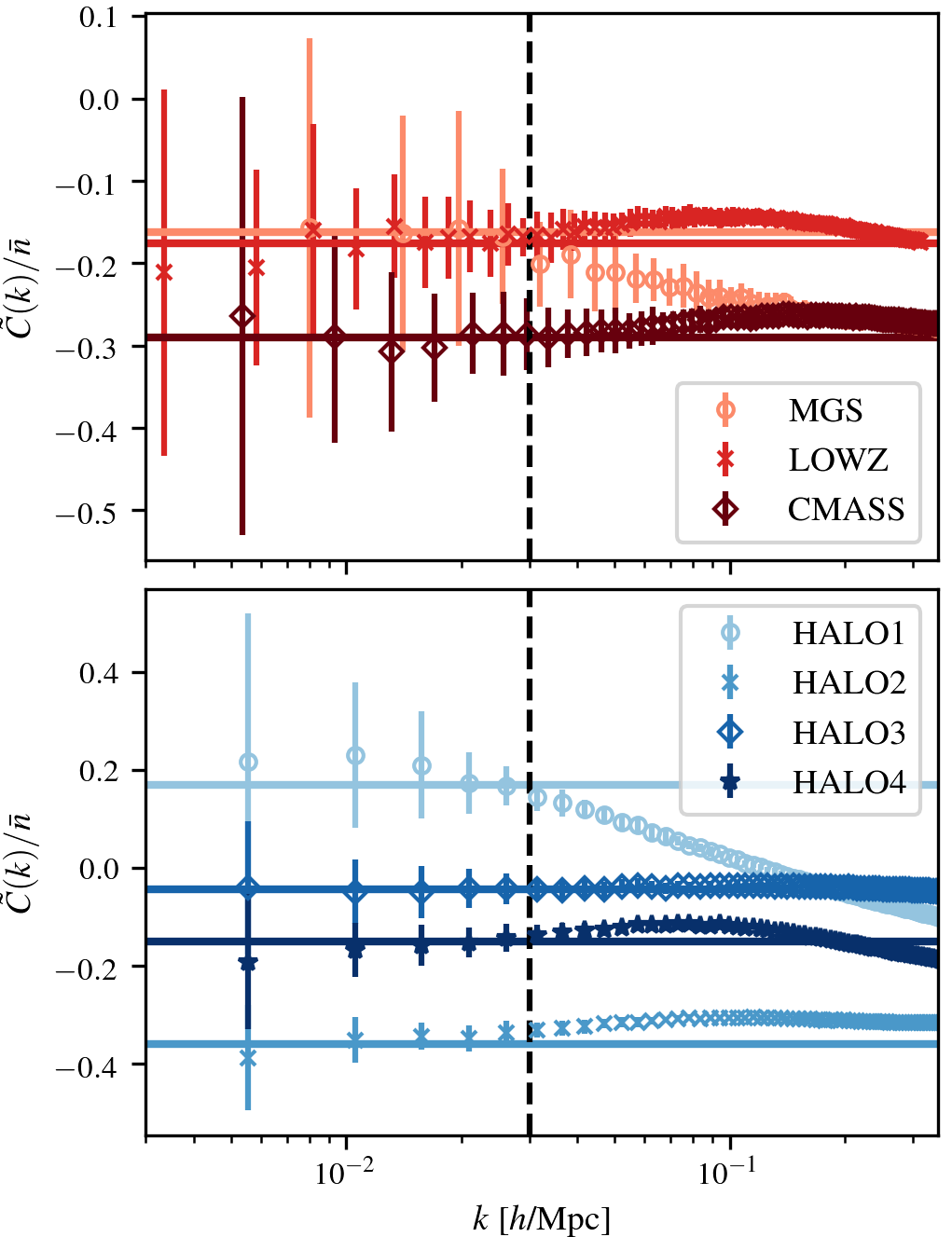}
  \caption{Measurements of the stochasticity power spectrum in units of Poisson shot noise for each
    sample. Solid lines represent large-scale fits to the data in order to determine $N_0$, while the vertical
    dashed line indicates the approximate cutoff scale for these fits (see text for precise numbers used in each
    case).}
  \label{fig:N0measurements}
\end{figure}

Using these measurements of the linear bias parameter we can construct the field
$\tilde{\varepsilon}_g(\B{k}) = \delta_g(\B{k}) - b_1\,\delta(\B{k})$, which corresponds to the stochastic field
introduced in Section~\ref{sec:stochasticity} on large scales, where contributions from higher-order bias terms are
negligible. The power spectrum of $\tilde{\varepsilon}_g$ can be expressed in terms of the measured galaxy and
matter power spectra \cite{HamSelDes1008,BalSelSmi1310}\,,
\begin{equation}
  \label{eq:data.Pstoch}
  \tilde{C}_{gg}(k) = P_{gg}(k) - 2b_1\,P_{gm}(k) + b_1^2\,P_{mm}(k)
\end{equation}
and should asymptote to Eq.~(\ref{eq:theory.Pstoch}) in the large-scale limit. This is demonstrated by
Fig.~\ref{fig:N0measurements}, which plots $\tilde{C}_{gg}$ in units of the Poisson noise of each sample and
shows that the stochasticity power spectrum approaches a constant on large scales, as expected. Moreover, we see
that this constant is negative for all but the low-mass halo sample at redshift $z = 0$, indicating that they
have sub-Poissonian noise and are thus dominated by the halo exclusion effect \cite{MoWhi9609,SheLem99,SmiScoShe0703,BalSelSmi1310}. We measure
the value of $N_0$ by fitting a constant to the data in Fig.~\ref{fig:N0measurements} using the same cutoff
scales as for the measurement of $b_1$ (approximately illustrated by the vertical dashed line). The results are
given in Table~\ref{tab:tracers} and are also shown by the solid lines. Unfortunately, it is not possible to
determine the scale-dependent component of the stochasticity directly from this data, as its effect is conflated
with higher-order bias terms to produce the deviations from a constant $\tilde{C}_{gg}$ for scales beyond $\sim
0.03\,\iMpc$.

In principle, the procedure followed here can be extended to measure higher-order bias parameters as shown in
\cite{LazSch1712,AbiBal1807}. While this could be an interesting consistency check with the results from the
model fits, we leave this possibility for future work.

\begin{table*}
  \centering
  \caption{Adopted prior bounds for all model parameters. All prior distributions are flat with the exception of
    $\gamma_2$ for which we choose a Gaussian with mean $\gamma_{2,\mathrm{ex}}(b_1)$ and standard deviation
    0.5. For the higher-derivative and scale-dependent noise parameters we use the arbitrary normalization scale
    $k_{\mathrm{HD}} = 0.4\,\iMpc$.}
  \begin{ruledtabular}
    \begin{tabular}{cccccccc}
      Catalog & $b_1$ & $b_2$ & $\gamma_2$ & $\gamma_{21}$ & $N_0$ [$1/\bar{n}$] &
       $\beta$, $\beta^{\times}$ [$k_{\mathrm{HD}}^{-2}$] & $N_2$, $N_2^{\times}$ [$k_{\mathrm{HD}}^{-2}/\bar{n}$] \Tstrut\Bstrut\\ 
      \hline
      MGS & [0.5, 3] & [-7, 7] & $G\left[\gamma_{2,\mathrm{ex}}(b_1),\,0.5\right]$ & [-7, 7] & [-1, 0.5] &
            [-50,50] & [-50, 50] \Tstrut \\ 
      LOWZ & [1, 4] & \dittotikz & \dittotikz & \dittotikz & \dittotikz & \dittotikz & \dittotikz \\
      CMASS & [1, 4] & \dittotikz & \dittotikz & \dittotikz & \dittotikz & \dittotikz & \dittotikz \\
      HALO1 & [0.5, 3] & \dittotikz & \dittotikz & \dittotikz & \dittotikz & \dittotikz & \dittotikz \\ 
      HALO2 & [1.5, 4.5] & \dittotikz & \dittotikz & \dittotikz & \dittotikz & \dittotikz & \dittotikz \\
      HALO3 & [1.7, 3.7] & \dittotikz & \dittotikz & \dittotikz & \dittotikz & \dittotikz & \dittotikz \\
      HALO4 & [2.5, 6.5] & [0,10] & \dittotikz & \dittotikz & \dittotikz & \dittotikz & \dittotikz \Bstrut \\ 
    \end{tabular}
  \end{ruledtabular}
  \label{tab:priors}
\end{table*}

\subsection{Fitting procedure and prior choices}
\label{sec:likelihood}

We now proceed to the description of how we infer model parameters from the measured power spectra. Firstly,
since the different realizations are statistically independent, we define the overall likelihood as the product
of all likelihood functions for a single ensemble member, which is just saying that we are simultaneously
fitting a given number of realizations with the same model, see~\cite{OddSefPor2003} for an identical
approach. We assume that the likelihood function for a single realization is given by a
multivariate Gaussian, so that
\begin{equation}
  \label{eq:data.likelihood}
  \begin{split}
    -2 \log{{\cal L}_{\mathrm{tot}}} &= -\frac{2}{N_R} \sum_{n=1}^{N_R} \log{{\cal L}_{(n)}} =
    \frac{1}{N_R}\sum_{n=1}^{N_R} \chi_{(n)}^2 \\ &\hspace{-1.5em}= \frac{1}{N_R}\sum_{n=1}^{N_R}
    \sum_{i,j=1}^{N_{\mathrm{bin}}} \left(X_i^{(n)} - \mu_i\right)\,C_{X,ij}^{-1}\,\left(X_j^{(n)} -
      \mu_j\right)\,,
  \end{split}
\end{equation}
where $X_i^{(n)}$ is the measurement from the $n$-th realization in bin $i$, $\mu_i$ the corresponding model
prediction and $C_{ij}$ the rescaled covariance matrix as described in Section~\ref{sec:measurements_P_cov}. The
factor $1/N_R$ ensures that after combining $N_R$ likelihood functions the sampling volume still corresponds to
our desired volume of $V_{\mathrm{eff,LOWZ}}$ and one can show that Eq.~(\ref{eq:data.likelihood}) is equivalent
(up to a constant) to the likelihood for the mean of the data with covariance $C_{X,ij}$. However, for all cases
considered here our sampling volume is smaller than the combined volume of the $N_R$ simulation boxes, so the
scatter in the data is less than statistically expected. We have checked that this has no significant impact on
the parameter posteriors, but has to be accounted for when using the $\chi^2$ as an indicator for the
goodness-of-fit (see Section~\ref{sec:goodness-if-fit}).

In a next step we minimize Eq.~(\ref{eq:data.likelihood}) by varying the model parameters using Markov chain
Monte Carlo (MCMC) and a standard Metropolis-Hastings sampling algorithm, leaving the cosmological parameters
fixed to their fiducial values. We generally adopt wide and flat prior distributions for all other parameters
and the exact bounds used in our analysis are given in Table~\ref{tab:priors}. The bounds of the $N_0$ prior are
determined from the condition that the overall stochasticity contribution to the power spectrum cannot be
negative and that super-Poisson noise values ($N_0 > 0$), if they occur, tend to be small (see
e.g. \cite{CasMoShe0207}). For the HALO4 sample we adapted the $b_2$ prior such that it is constrained to
positive values in order to prevent a strongly bimodal posterior distribution. This behavior is caused by the
$b_2^2$ term in Eq.~(\ref{eq:theory.Pgg_1loop}), which comes to dominate on small scales because the high
redshift and the highly biased nature of this particular sample lead to a generically large $b_2$. Since the
peak-background split prediction \cite{LazWagBal1602} strongly favors a positive $b_2$ given the linear bias
parameter of this sample, we choose the prior to be positive. The only exception in our list of priors is the
tidal bias parameter $\gamma_2$, which we assume to be a Gaussian centered on the excursion set relation
$\gamma_{2,\mathrm{ex}}(b_1)$ (see Eq.~\ref{eq:theory.g2ECS}) with standard deviation 0.5. We stress that the
mean of this prior depends on the linear bias parameter and therefore changes in each step of the Markov
chain. This choice is motivated by 1) an otherwise strong degeneracy\footnote{This degeneracy is caused by the
  the $\gamma_2$ and $\gamma_{21}$ contributions to the one-point propagator (Eq.~\ref{eq:theory.MP1}), which
  can be shown to be \emph{exactly} degenerate.} between $\gamma_2$ and $\gamma_{21}$, and 2) the good agreement
(much better than the width of the prior) between the excursion set relation and independent measurements of
$\gamma_2$ from halos.

For every fit that we perform we run several independent Markov chains and determine their convergence by means
of the Gelman-Rubin criterion \cite{GelRub9201}, specifically $R - 1 < 0.01$, but make sure that they run for at
least a total of 120,000 accepted steps. After removing the burn-in these are combined into a single chain,
which we pass to the software package \texttt{getdist} \cite{Lew2019} to extract the parameter posteriors.

\subsection{Performance metrics}
\label{sec:metrics}

Adding complexity to the model of the galaxy power spectrum can increase its range of validity down to smaller
scales when compared to data, but the price to pay is a larger set of nuisance parameters that have to be
marginalized over in order to arrive at the desired constraints on any cosmological parameters. Vice versa the
application of co-evolution relations as discussed in Section~\ref{sec:co-evolution} can improve the merit of the
model, but potentially only over a rather limited range of scales. Clearly, there is a compromise to be found
between the validity and merit of our perturbative models. To quantify such a balance we compute various
indicators from our Markov chains, following a similar analysis presented in \cite{OsaNisBer1903} that focused
on the matter power spectrum alone.

\subsubsection{Figure of bias}
\label{sec:fob}

One quality of a robust model must be its ability to return unbiased estimates of model parameters. Having
determined the posterior means (symbolized by an over-bar) of a set of parameters $\theta_{\alpha}$ along with
their covariance matrix $S_{\alpha\beta}$, we can define the following simple measure, which we denote as the
\emph{Figure of Bias} (FoB):
\begin{equation}
  \label{eq:data.FoB}
  \mathrm{FoB} \equiv \left[\sum_{\alpha,\beta}
    \left(\bar{\theta}_{\alpha}-\theta_{\mathrm{fid},\alpha}\right)\,\tilde{S}^{-1}_{\alpha\beta}
    \left(\bar{\theta}_{\beta}-\theta_{\mathrm{fid},\beta}\right)\right]^{1/2}\,.
\end{equation}
Here we have additionally taken into account any uncertainty in the fiducial parameters by writing
$\tilde{S}_{\alpha\beta} = S_{\alpha\beta} + S_{\mathrm{fid},\alpha\beta}$. Since in this work we do not vary
the cosmological parameters, we define the FoB solely with respect to the linear bias parameter and consider the
measurements detailed in Section~\ref{sec:measurements_bias_noise} as the truth, in which case
Eq.~(\ref{eq:data.FoB}) simplifies to
$\mathrm{FoB} = \left(\bar{b}_1 - b_{1,\mathrm{fid}}\right)/\sqrt{\sigma^2_{b_1}+\sigma^2_{\mathrm{fid},b_1}}$
with $\sigma_{\mathrm{fid},b_1}$ denoting the uncertainties reported in Table~\ref{tab:tracers}. As an amplitude
parameter, which scales up and down contributions from different terms to the power spectrum, it can be regarded
as a proxy for $\sigma_8$. However, we caution that it is most likely not representative of other cosmological
parameters, such as $\Omega_m$ and $h$, which leave a stronger impact on the baryon acoustic oscillation
signature and its overall shape. As a test of the galaxy bias modeling our definition of the FoB should
certainly be adequate.

\subsubsection{Goodness-of-fit}
\label{sec:goodness-if-fit}

The FoB alone is insufficient to judge the validity of a model. It is easy to imagine a situation, particularly
so when the FoB is only based on a subset of all model parameters, where one recovers the fiducial values, while
the model is actually not a good description of the data. For that reason we also need to assess the
goodness-of-fit, which we quantify in terms of the minimum standard $\chi^2$ values that are computed as part of
our likelihood, i.e., $\chi^2_{\mathrm{tot}} = 1/N_R \sum_{n=1}^{N_R}\chi^2_{(n)}$
(see. Eq.~\ref{eq:data.likelihood}). However, as already mentioned in Section~\ref{sec:likelihood}, when
considering the total amount of data the measurement uncertainties used in our analysis are statistically too
large by a factor of $N_R/\eta$, so we have to rescale the $\chi^2$ accordingly to get a meaningful value:
\begin{equation}
  \label{eq:data.chi2rescaling}
  \tilde{\chi}^2_{\mathrm{tot}} = \frac{N_R}{\eta}\,\chi^2_{\mathrm{tot}}\,.
\end{equation}
This value can subsequently be compared to the $68\,\%$ or $95\,\%$ confidence limits of the
$\chi^2$-distribution with $\mathrm{dof}$ degrees of freedom to determine the goodness-of-fit. The degrees of
freedom are given by
\begin{equation}
  \label{eq:data.dof}
  \mathrm{dof} = N_R \times N_{\mathrm{bin}} - N_p\,,
\end{equation}
where $N_{\mathrm{bin}}$ and $N_p$ are the number of bins and model parameters, respectively.

In the following we will also perform MCMC runs where we replace the nonlinear matter power spectrum (i.e., the
term that gets multiplied by $b_1^2$ or $b_1$ in the galaxy auto and cross power spectrum) with direct
measurements from the underlying matter fields. In that case we inadvertently model part of the scatter in the
data, which means that the difference between data $X_i = P_{gg}(k_i)$, $P_{gm}(k_i)$ and model $\mu_{X,i}$ is
no longer given by a multivariate Gaussian with covariance matrix $C_{X,ij}$. Writing
$\Delta_i = X_i - b_1^{n_X}\,P_{mm}(k_i)$ and $\mu_{\Delta,i} = \mu_{X,i} - b_1^{n_X}\,\mu_{P_{mm},i}$ we can
show instead that $\Delta_i - \mu_{\Delta,i}$ obeys the modified covariance matrix
\begin{equation}
  \begin{split}
      C_{\Delta,ij} &= \left<\left(\Delta_i-\mu_{\Delta,i}\right)\,\left(\Delta_j-\mu_{\Delta,j}\right)\right> \\ &=
      C_{X,ij} + b_1^{2n_X}\,C_{m,ij} - 2b_1^{n_X}\,C_{X \times m,ij}\,,
  \end{split}
\end{equation}
where $n_X = 2\,(1)$ for $X = P_{gg}\,(P_{gm})$ and $C_{m,ij}$ and $C_{X \times m,ij}$ are the covariance of the
matter field and its cross-covariance with $X$. This leads to a reduction of the inferred $\chi^2$ value by the
amount
\begin{equation}
  \Delta\chi^2 = -b_1^{n_X} \sum_{i,j} \left(2 C_{X \times m,ij} - b_1^{n_X}\,C_{m,ij}\right)\,C_{X,ij}^{-1}\,,
\end{equation}
which we correct for using the fiducial value of $b_1$ before applying the rescaling according to
Eq.~(\ref{eq:data.chi2rescaling}). For joint analyses of the auto and cross power spectrum we proceed in the
analogous fashion.

\subsubsection{Figure of merit}
\label{sec:fom}

Finally, we define the constraining power of a given model by the reciprocal of the posterior volume
corresponding to the $68\,\%$ confidence limit. This is related to the determinant of the parameter covariance
matrix and so we write our \emph{Figure of Merit} (FoM) as
\begin{equation}
  \label{eq:data.FoM}
  \mathrm{FoM} \equiv \frac{1}{\sqrt{\mathrm{det}\,\left[S_{\alpha\beta}/\left(\theta_{\mathrm{fid},\alpha}\theta_{\mathrm{fid},\beta}\right)\right]}}\,.
\end{equation}
The inclusion of the additional factor $\theta_{\mathrm{fid},\alpha}\,\theta_{\mathrm{fid},\beta}$ ensures that
the FoM is defined in terms of the relative parameter uncertainties, which yields more comparable results for
our different samples. As for the FoB we focus on the linear bias parameter, such that we simply have
$\mathrm{FoM} = b_{1,\mathrm{fid}}/\sigma_{b_1}$.

\begin{figure*}
  \centering
  \includegraphics[width=\textwidth]{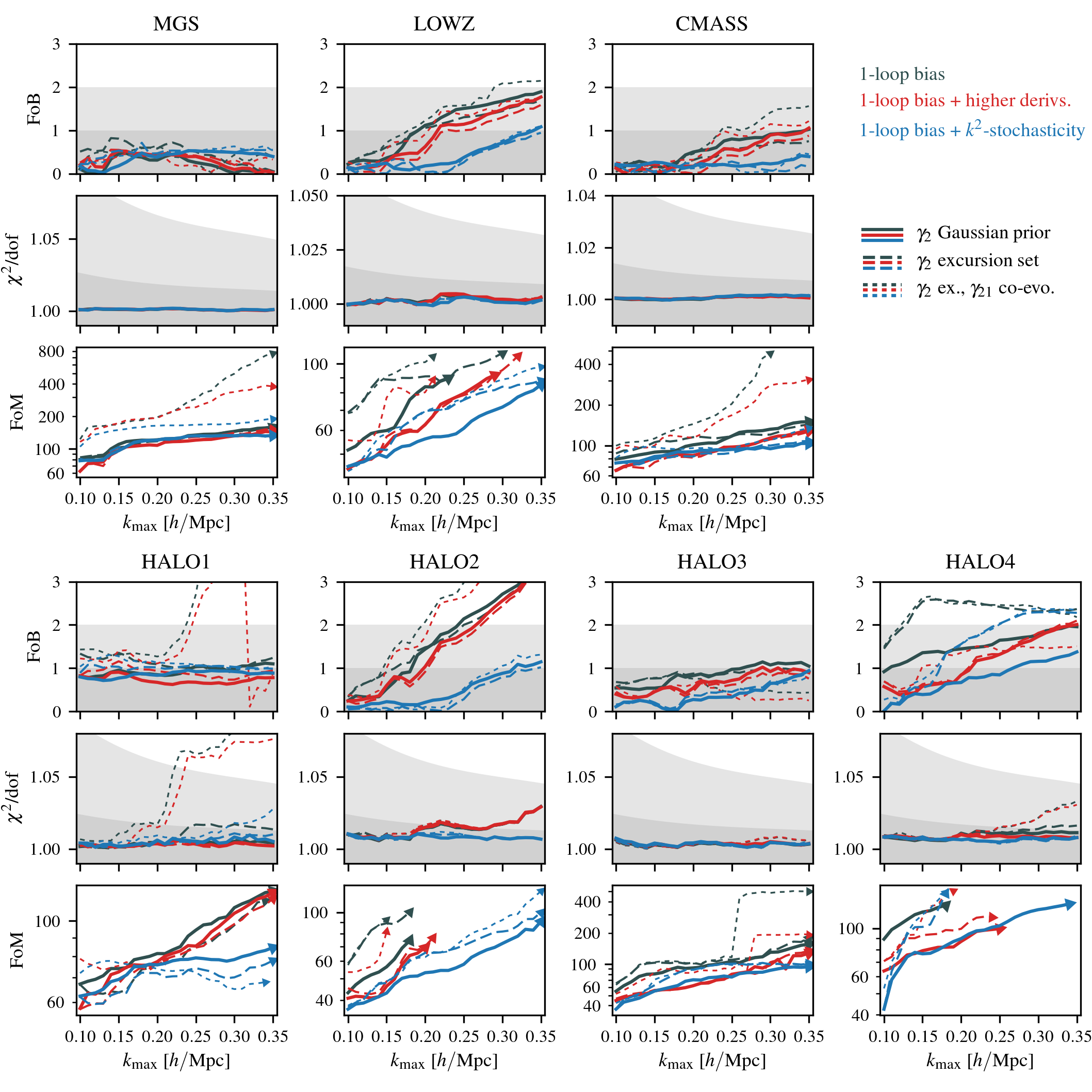}
  \caption{FoB, goodness-of-fit and FoM for the auto power spectrum as a function of the maximum $k$-mode
    included in the fit (note the differing $y$-axis ranges for the FoM). Grey shaded areas indicate the
    $68\,\%$ and $95\,\%$ confidence limits for the FoB and the $\chi^2$ distribution, which are used to assess
    the breakdown of the model, at which point we stop plotting the FoM. Colors distinguish between the standard
    one-loop bias model (black, 5 free parameters) and extensions to higher derivatives (red, 6 free param.) and
    scale-dependent noise (blue, 6 free param.). Thick, solid lines only impose a Gaussian prior centered on
    $\gamma_{2.\mathrm{ex}}(b_1)$, while the thinner dashed and dotted lines have $\gamma_2$, or both,
    $\gamma_2$ and $\gamma_{21}$ fixed in terms of the excursion set and co-evolution relations.}
  \label{fig:fcf_Pgg_trueM}
\end{figure*}

\section{Testing one-loop galaxy bias}
\label{sec:results_true_matter}

We now turn to the main goal of the paper, that is, to test the regime of validity of one-loop galaxy bias, and to see which effects in the bias expansion are most important. To do so \emph{we use the measured nonlinear matter spectrum} in place of $P_{mm}$, as this allows us to concentrate on bias independently of any issues related to the nonlinear evolution of matter description. As mentioned in the previous section, we test a variety of biased tracers at different redshifts to extract robust conclusions about which bias effects are generically important. Also note that by ignoring redshift-space distortions, we are  avoiding extra parameters that can mask failures of the bias model. Furthermore, we check that the bias parameters we obtain from our MCMC chains satisfy basic sanity checks with independent measurements we can make and/or fitting formulae when available in the literature. Our most stringent  test asks the bias model to simultaneously  match the auto (galaxy-galaxy) and cross (galaxy-mass) spectrum, but we start with the most common (weaker) test of using the auto spectrum alone.  In addition, we  investigate the relevance of the higher-derivative and scale-dependent noise terms, as well as the impact of using co-evolution relations to reduce the number of
free fitting parameters.

\subsection{Validity of one-loop galaxy bias for  the auto power spectrum}
\label{sec:validity-auto-power}

In this section we analyze the performance of the galaxy auto power spectrum for three different modeling
options: 1) only including terms from the one-loop galaxy bias expansion (also referred to as ``standard'' model
in the following), 2) taking also into account short-range nonlocality and the resulting higher-derivative
contribution, and 3) considering scale-dependent stochasticity instead of higher-derivative bias. While the
first option has five free parameters in total ($b_1$, $b_2$, $\gamma_2$, $\gamma_{21}$ and $N_0$), the other
two have one extra parameter each.

\subsubsection{Fiducial survey volume}
\label{sec:validity-Pgg.fidvol}

We plot the derived FoB, reduced minimum $\chi^2$ and FoM of these three
models in Fig.~\ref{fig:fcf_Pgg_trueM}, represented by the thick, solid lines. Our metrics are shown as a
function of the maximum mode $k_{\mathrm{max}}$ that is included in the fitting procedure up to
$k_{\mathrm{max}} = 0.35\,\iMpc$, but for ease of comparison we introduce a scale $k_{\dagger}$ at which point
the model is deemed to fail and we truncate the FoM, which is indicated by a triangle. We define this scale as
the combination of FoB and minimum $\chi^2$ reaching a certain critical value, specifically
\begin{equation}
  \label{eq:res1.kfail}
  \mathrm{FoB}(k_{\dagger}) + \frac{\chi^2(k_{\dagger}) -
    \mathrm{dof}(k_{\dagger})}{\chi^2_{95\,\%}(k_{\dagger})-\mathrm{dof}(k_{\dagger})} = \sigma_{\mathrm{crit}} 
\end{equation}
with $\sigma_{\mathrm{crit}} = 1.5$\footnote{The choice of the threshold for $\sigma_{\mathrm{crit}}$ is
  somewhat arbitrary, but we do vary it to make our threshold more stringent, which can alternatively be
  interpreted as scaling our results for larger survey volumes (see Eq.~\ref{eq:res1.sigmacrit} and
  Fig.~\ref{fig:fom_fail_trueM} in connection to this). }. This means we allow for a maximum bias of
$1.5\,\sigma$ if the minimum $\chi^2$ matches the degrees of freedom (i.e., the expected $\chi^2$ for a good fit
to the data), and similarly, if the fit is completely unbiased, the discrepancy between $\chi^2$ and dof can be
as large as one and a half times the corresponding value for the $95\,\%$ confidence limit. The gray shaded
areas in Fig.~\ref{fig:fcf_Pgg_trueM} mark the $68\,\%$ and $95\,\%$ limits for the FoB and $\chi^2$.

We first notice that the standard model (black) performs very well and delivers unbiased constraints on $b_1$ as
well as a good fit to the measurements for a broad range of scales. In fact, according to our criterion it does
not break down before $k_{\mathrm{max}} = 0.35\,\iMpc$ for all samples with the exception of the LOWZ and the
two high-mass halo samples (HALO2 and HALO4), for which the model stops working at
$k_{\mathrm{max}} \sim 0.2\,\iMpc$. In these cases we can extend its range of validity by including either a
higher-derivative (red) or scale-dependent noise term (blue) and whereas the former does not yield significantly
smaller FoBs, the latter allows us to fit the data again up to $k_{\mathrm{max}} = 0.35\,\iMpc$. However, as
anticipated, this comes with a penalty in the FoM and comparing the maximally achievable FoM for LOWZ, HALO1 and
HALO2 the standard and scale-dependent noise models are about equal. In all other samples adding a
scale-dependent noise parameter has a bigger impact on the FoM than the higher-derivative term and leads to a
decrease of $\sim 25\,\%$ in the FoM (at least within the tested range of scales).

Furthermore, we study how our performance metrics change when fixing $\gamma_2$ using the excursion set relation
from Eq.~(\ref{eq:theory.g2ECS}) and also by additionally constraining $\gamma_{21}$ to co-evolution as a
function of both $b_1$ and $\gamma_2$ (see Eq.~\ref{eq:theory.g21evo}). These cases are shown by the dashed and
dotted lines in Fig.~\ref{fig:fom_fail_trueM} for each of the three modeling options. Focusing on the dashed
lines first, we observe that they tend to be valid over the same range of scales as when $\gamma_2$ is being
varied under a Gaussian prior with the exception of HALO4, where fixing $\gamma_2$ gives rise to a significant
increase in FoB. Moreover, they yield improvements of the FoM that are of the order $20$-$30\,\%$ for LOWZ and
HALO2, but more modest in all other cases (some even have lower FoM), which seems to imply a dominance of the
Gaussian prior. Remarkably, fixing $\gamma_2$ and $\gamma_{21}$ at the same time does not generally lead to any
stricter limitations in the range of validity apart from the HALO1 sample, whose FoB and minimum $\chi^2$ rise
quickly for $k_{\mathrm{max}} \gtrsim 0.2\,\iMpc$. However, the benefit can be much higher and in particular for
the standard model can result in improvements that are as large as a factor of three (e.g. MGS and CMASS).

\begin{figure}
  \centering
  \includegraphics[width=\columnwidth]{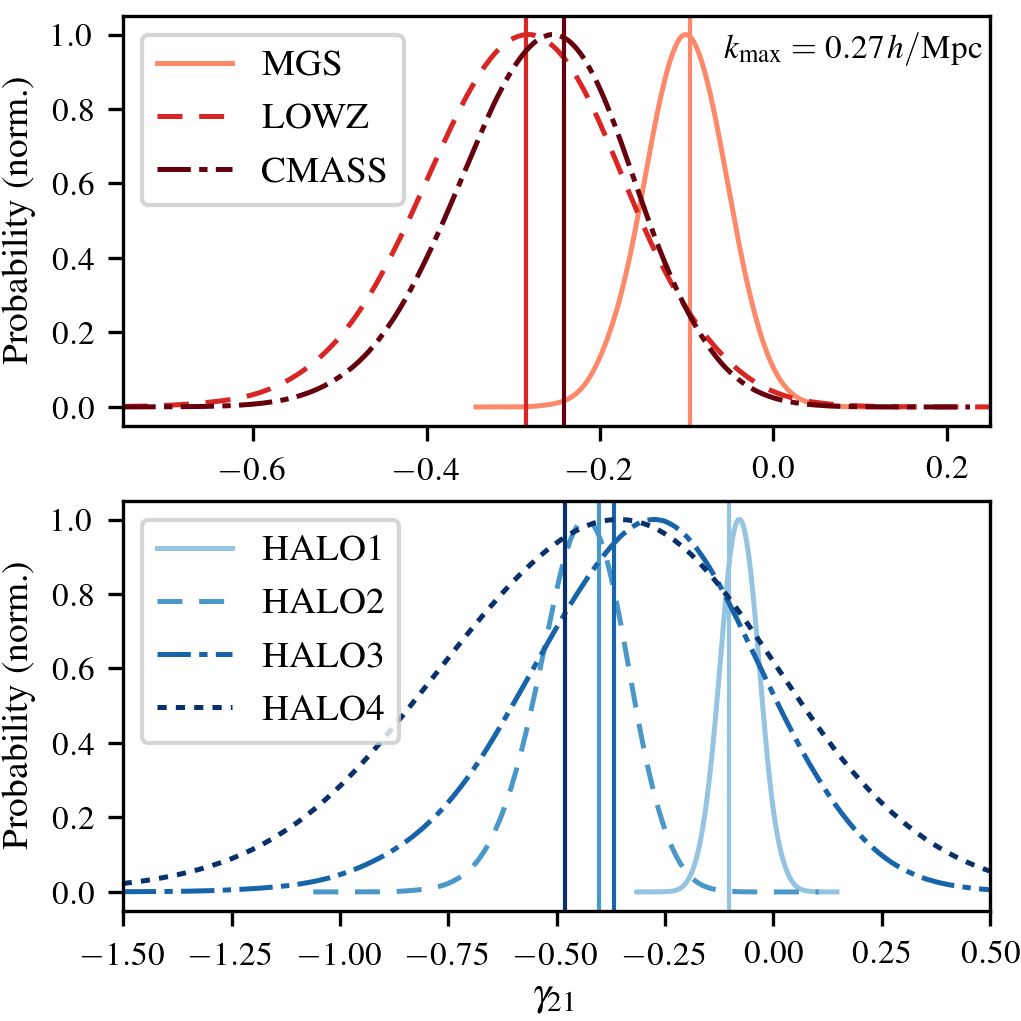}
  \caption{1d marginalized posteriors of $\gamma_{21}$ from the auto power spectrum at
    $k_{\mathrm{max}} = 0.27\,\iMpc$ with scale-dependent noise and $\gamma_2$ fixed to the excursion set
    relation. The vertical lines of matching colors correspond to the co-evolution relation for $\gamma_{21}$
    for each of the samples.}
  \label{fig:g21_from_Pgg}
\end{figure}

We also note that for many samples $\gamma_{21}$ is clearly constrained to be non-zero, even for moderate scales
$k_{\mathrm{max}} \sim 0.2\,\iMpc$, when we fix $\gamma_2$ and thus break their degeneracy. This can be seen in
Fig.~\ref{fig:g21_from_Pgg}, where we plot the fully marginalized posterior distribution of $\gamma_{21}$ for
the exemplary case of the scale-dependent noise model at $k_{\mathrm{max}} = 0.27\,\iMpc$. In addition, the
figure shows that the $\gamma_{21}$ constraints are typically in good agreement with the prediction from the
co-evolution relation, which explains the good model performances observed in Fig.~\ref{fig:fcf_Pgg_trueM}. We
therefore conclude that ignoring $\gamma_{21}$ in models of the galaxy power spectrum is highly disfavored
opposed to employing the co-evolution relation.

\subsubsection{Estimating the dependence on survey volume}
\label{sec:validity-Pgg.volume}

\begin{figure*}
  \centering
  \includegraphics[width=\textwidth]{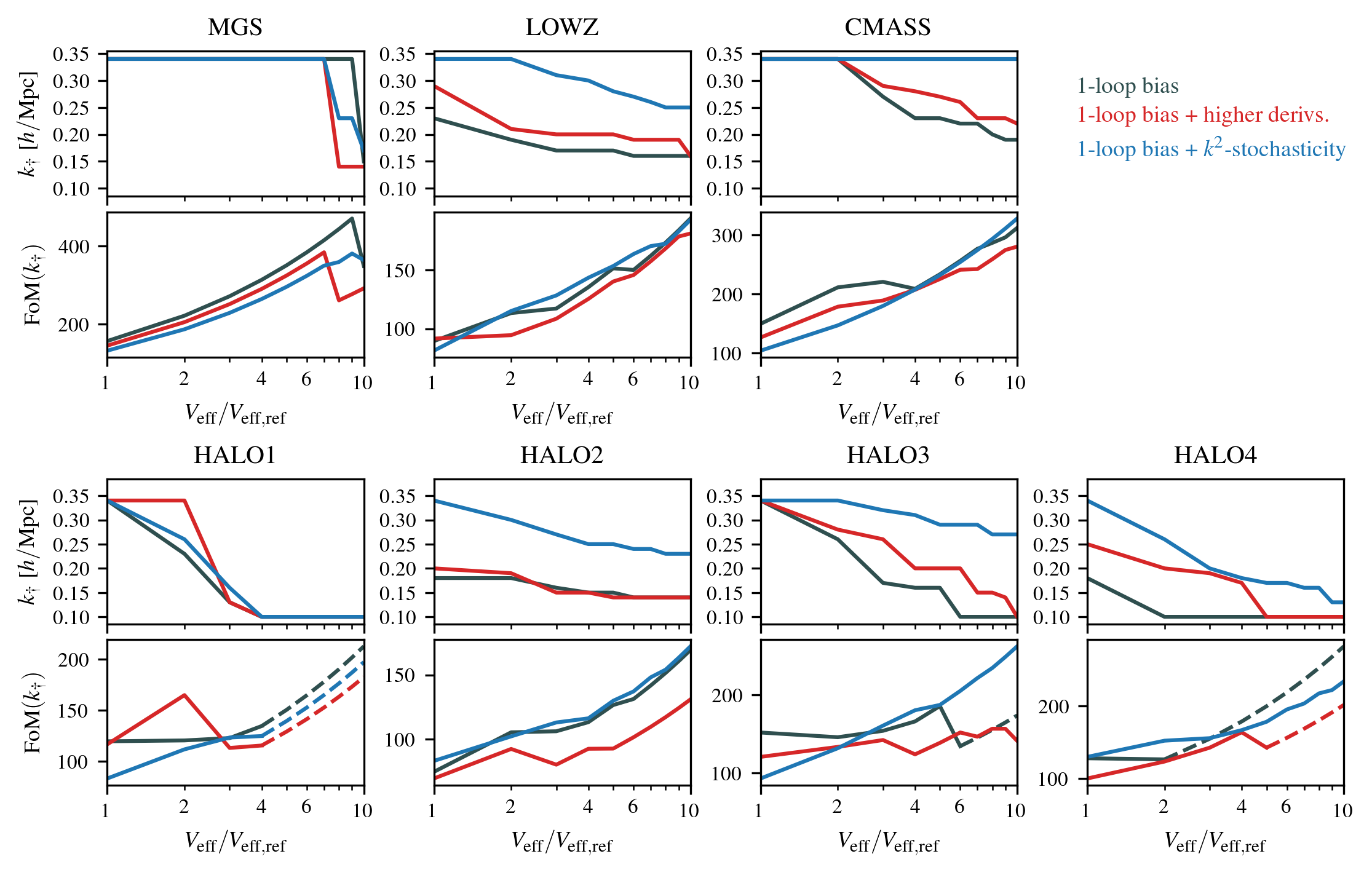}
  \caption{Maximum scale of validity, $k_{\dagger}$, and corresponding FoM (scaled up by
    $\sqrt{V_{\mathrm{eff}}/V_{\mathrm{eff,ref}}}$) as a function of volume relative to our reference
    $V_{\mathrm{eff,ref}} = V_{\mathrm{eff,LOWZ}} \approx 6\,(\Gpc)^3$. Alternatively, one can regard this as a
    function of an increasingly restrictive validity criterion (see Eq.~\ref{eq:res1.sigmacrit}). Different
    colors represent different modeling options with a Gaussian prior on $\gamma_2$. If the model would break
    down before $k_{\mathrm{max}} = 0.1\,\iMpc$, the FoM is evaluated at $k_{\dagger} = 0.1\,\iMpc$, which is
    indicated by the dashed lines.}
  \label{fig:fom_fail_trueM}
\end{figure*}

We can now raise the question how these results depend on our criterion that defines the validity of the model
and the adopted survey volume. To a first approximation these two questions are equivalent as a simple estimate
reveals: increasing the volume by a factor $V_{\mathrm{eff}}/V_{\mathrm{eff,ref}}$ ($V_{\mathrm{eff,ref}}$
denotes our reference volume of $V_{\mathrm{eff,LOWZ}}$) means that the parameter uncertainties decrease by the
square root of that factor and hence the FoB increases accordingly. While the $\chi^2$ grows linearly with
volume, so do the degrees of freedom, such that the ratio $\chi^2/\mathrm{dof}$ must stay invariant. However,
the $95\,\%$ confidence limit behaves in a different way, namely
$\chi^2_{95\,\%}/\mathrm{dof} - 1 \propto 1/\sqrt{\mathrm{dof}}$, which can be easily shown from the fact that
the $\chi^2$ distribution is well approximated by a Gaussian with mean $\mathrm{dof}$ and standard deviation
$\sqrt{2\mathrm{dof}}$ for $\mathrm{dof} \gg 1$. Therefore, the second term in Eq.~(\ref{eq:res1.kfail}) also
scales as the square root of the volume factor and we can write
\begin{equation}
  \label{eq:res1.sigmacrit}
  \sigma_{\mathrm{crit}}(V_{\mathrm{eff}}) =
  \sigma_{\mathrm{crit}}(V_{\mathrm{eff,ref}})\,\sqrt{\frac{V_{\mathrm{eff,ref}}}{V_{\mathrm{eff}}}}\,. 
\end{equation}
In reality this scaling will not be exactly satisfied because of parameter degeneracies and noise in the data,
but it can be exploited to glean useful insights into how our results extrapolate to larger survey volumes.

Using Eq.~(\ref{eq:res1.sigmacrit}) we solve Eq.~(\ref{eq:res1.kfail}) for $k_{\dagger}$ as a function of volume
and determine the corresponding FoM, which we scale up by $\sqrt{V_{\mathrm{eff}}/V_{\mathrm{eff,ref}}}$. The
results, ranging from our nominal volume up to a ten-fold increase, are shown in Fig.~\ref{fig:fom_fail_trueM}
for the standard, higher-derivative and scale-dependent noise models when only the Gaussian prior on $\gamma_2$
is applied. Note that we are limited in our $k_{\mathrm{max}}$ range from $0.1\,\iMpc$ to $0.35\,\iMpc$, so if
the validity criterion is not met even for the smallest mode, we opt to compute the FoM for
$k_{\dagger} = 0.1\,\iMpc$, which is clearly an over-estimation and thus marked by dashed lines in the
figure. 

As expected, we see that an increase in volume leads to a decrease in the range of validity for all three models
with the extension to scale-dependent stochasticity consistently providing the largest $k_{\dagger}$ values,
followed by the higher-derivative and standard model. In particular, even for a ten-fold increase in volume
the former still proves to be robust up to $~ 0.2\,\iMpc$ and in several cases beyond. A special case is the
HALO1 sample, which quickly reaches the lower limit for $k_{\dagger}$, caused by the FoB already being of order
one for our fiducial volume. Since this is independent of $k_{\mathrm{max}}$ (see Fig.~\ref{fig:fcf_Pgg_trueM}) and
we find smaller FoBs when combining with the cross power spectrum in Section~\ref{sec:consistency-ax}, this can
likely be attributed to a projection effect when marginalizing over the posterior. 

Furthermore, Fig.~\ref{fig:fom_fail_trueM} demonstrates that the decrease in $k_{\dagger}$ follows roughly a
power law, whose slope is similar for the various models, but varies from sample to sample. However, for CMASS
and HALO3 it is notably shallower when allowing for scale-dependent stochasticity, which has important
consequences for its FoM, as the initial discrepancy compared to the standard model can be compensated at larger
volumes. More generally we see that with increasing volume the scale-dependent noise model always gives rise to
either equal or better FoMs than the standard model and so its extended $k_{\mathrm{max}}$ range can overcome
the penalty of having an extra free parameter --- on the other hand, this is not true for the higher-derivative
model.

\subsection{Consistency between auto and cross power spectra}
\label{sec:consistency-ax}

\begin{figure*}
  \centering
  \includegraphics[width=\textwidth]{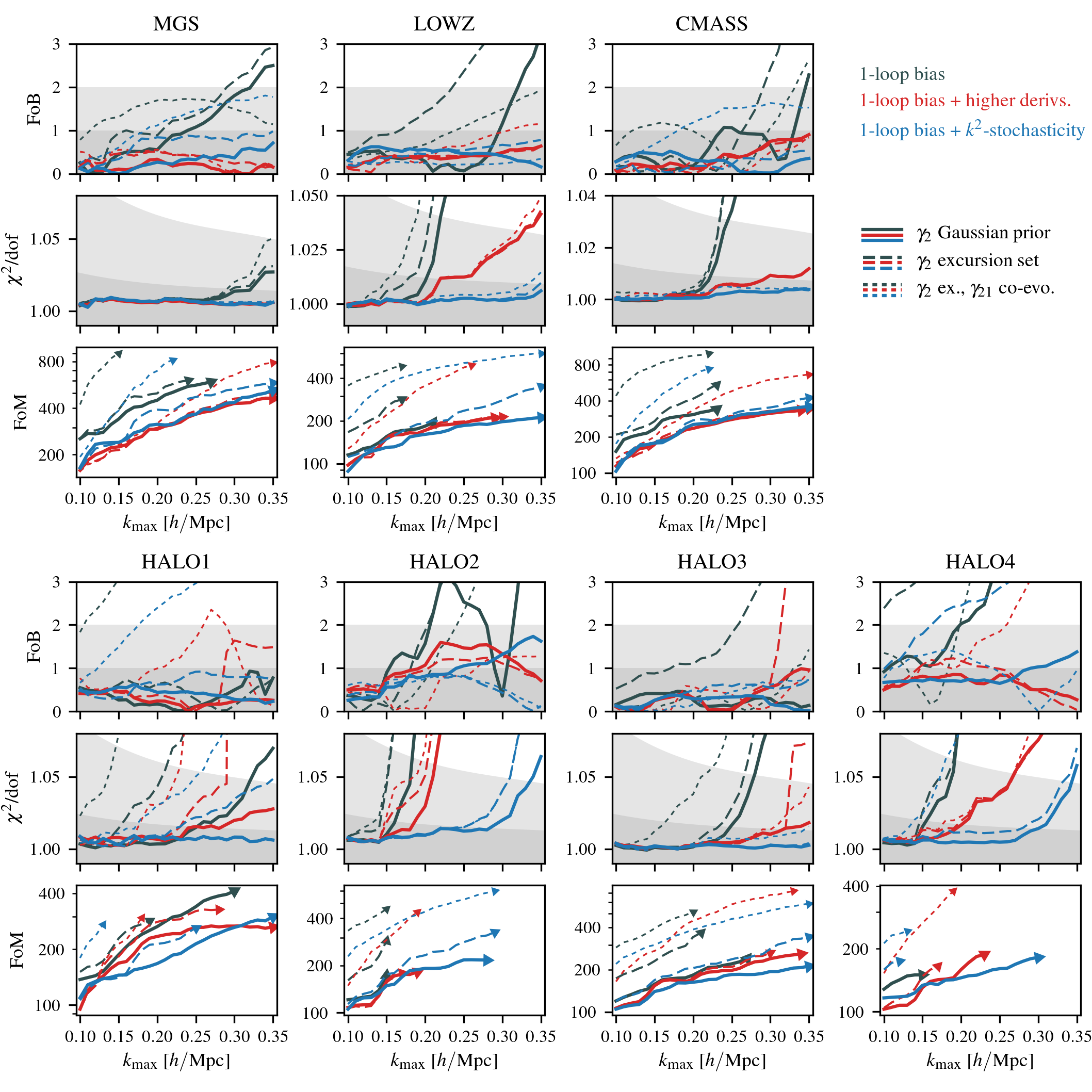}
  \caption{Same as Fig.~\ref{fig:fcf_Pgg_trueM}, but for the combination of the auto and cross power
    spectrum. Note that the number of free model parameters are 5 (black) and 7 (red/blue) for the case when
    only a Gaussian prior on $\gamma_2$ is employed.}
  \label{fig:fcf_Pgg+Pgm_trueM}
\end{figure*}

We now move on to check over which range of scales the various models already discussed in the previous section
are able to make consistent predictions for the auto and cross power spectra. When dealing with a single
observable whose model involves a large enough parameter space, it is possible that a failure of the model can
be disguised by the nuisance parameters (meaning here parameters whose values are not being tracked by the FoB)
absorbing any lacking contributions. Consistency in the auto and cross power spectra is therefore a much more
stringent test of the one-loop galaxy bias model, especially because from a perturbation theory point of view one
would expect these two statistics to be valid over the same range of scales.

\begin{figure*}
  \centering
  \includegraphics[width=\textwidth]{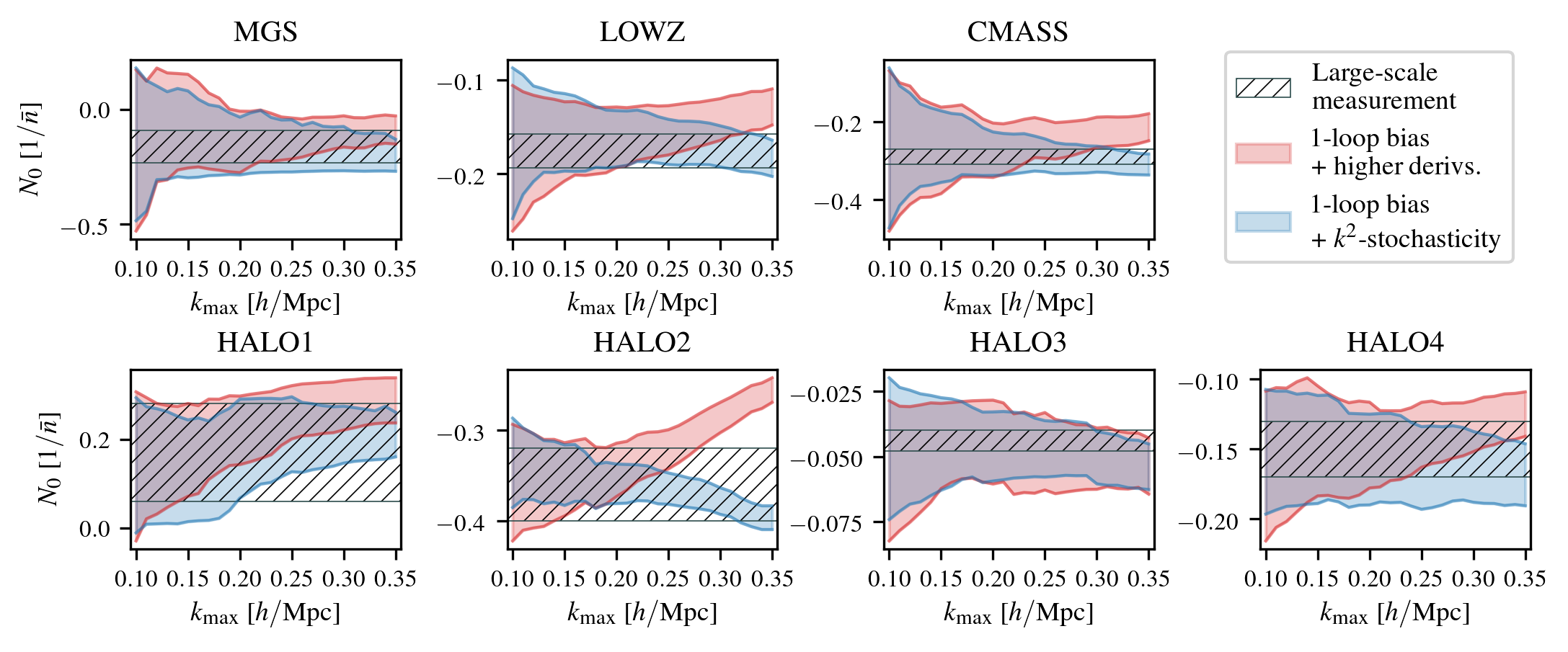}
  \caption{Constraints on large-scale deviations from Poisson shot noise, as captured by the parameter
    $N_0$. The red and blue shaded error bands show the 1-$\sigma$ uncertainties obtained from a joint fit of
    the auto and cross power spectrum at a given $k_{\mathrm{max}}$ including either a higher-derivative or
    scale-dependent noise term. The hatched area represents the constraint from the large-scale
    (model-independent) fit as described in Section~\ref{sec:measurements_bias_noise}.}
  \label{fig:N0_trueM}
\end{figure*}

In Fig.~\ref{fig:fcf_Pgg+Pgm_trueM} we present the three performance metrics derived from jointly analyzing the
measurements of the auto and cross power spectra. While the standard model still only has five free parameters,
the other two now have seven each as we allow for independent higher-derivative or scale-dependent stochasticity
effects in $P_{gg}$ and $P_{gm}$. Focusing to begin with on the thick, solid lines again, we observe that the
range of validity of the standard model is reduced by up to $30\,\%$ compared to the auto power spectrum
alone. This can be attributed to a quick increase in the minimum reduced $\chi^2$, which happens for most
samples at scales $k_{\mathrm{max}} \sim 0.2\,\iMpc$ (somewhat later for MGS and HALO1) and indicates an arising
inconsistency between the auto and cross power spectra. Indeed, by comparing the parameter posteriors of the
individual fits, we find that the scale at which the rise in $\chi^2$ occurs is typically accompanied by a
mismatch of marginalized parameter constraints, particularly for the second-order bias $b_2$.

Introducing higher-derivative terms helps alleviating this inconsistency as the decrease in the $\chi^2$ values
shows, but it does not resolve it, which is especially evident for LOWZ and the two high-mass halo samples. On
the other hand, the scale-dependent noise model brings about significant improvements for all samples without
any indication of breaking down in the range of scales probed, except for HALO2 and HALO4, where it remains
valid until $k_{\mathrm{max}} \sim 0.3\,\iMpc$. This suggests that for our most massive objects either both,
higher-derivative and scale-dependent noise effects, eventually become relevant, or the lack of two-loop terms
from the general bias expansion. Comparing the various FoMs we come to similar conclusions as for the auto power
spectrum alone: in cases where the standard model fails early on, the extended range of the more complex models
can compensate for their extra free parameters, but does not lead to significantly larger FoMs. As discussed
above, this might change when a larger survey volume (or stricter validity requirement) is considered.

As for the auto power spectrum alone, using the excursion set relation for $\gamma_2$ does generally not
diminish any of the three model's range of validity (as before with the exception of HALO4 and to a lesser
degree HALO1). However, since the combination of $P_{gg}$ and $P_{gm}$ reduces the impact of the Gaussian prior
on the $\gamma_2$ posterior, the improvements in FoM are consistently larger and of the order $\sim 10$ -
$20\,\%$ in most cases. Further and even more substantial improvements can be achieved by fixing $\gamma_{21}$
to its co-evolution relation, although we now observe that this leads to premature failures of the model for
more samples than in case of $P_{gg}$.

Finally, an interesting feature of the FoM graphs for various modeling options in
Fig.~\ref{fig:fcf_Pgg+Pgm_trueM} is their tendency to flatten off towards large $k_{\mathrm{max}}$. This
suggests that the information that can be extracted from the nonlinear regime, at least for the linear bias
parameter, is saturated for the combination of the auto and cross power spectrum beyond a certain value of
$k_{\mathrm{max}}$.

\subsection{Constraints on stochasticity and higher-derivative parameters}
\label{sec:noise-HD-constraints}

In the previous two sections we have seen that the scale-dependent noise model provides a more accurate and less
biased description of the measurements in the nonlinear regime than the extension to higher-derivatives. We now
intend to further investigate this assertion by considering the derived constraints on the stochasticity and
higher-derivative parameters.

\begin{figure*}
  \centering
  \includegraphics[width=\textwidth]{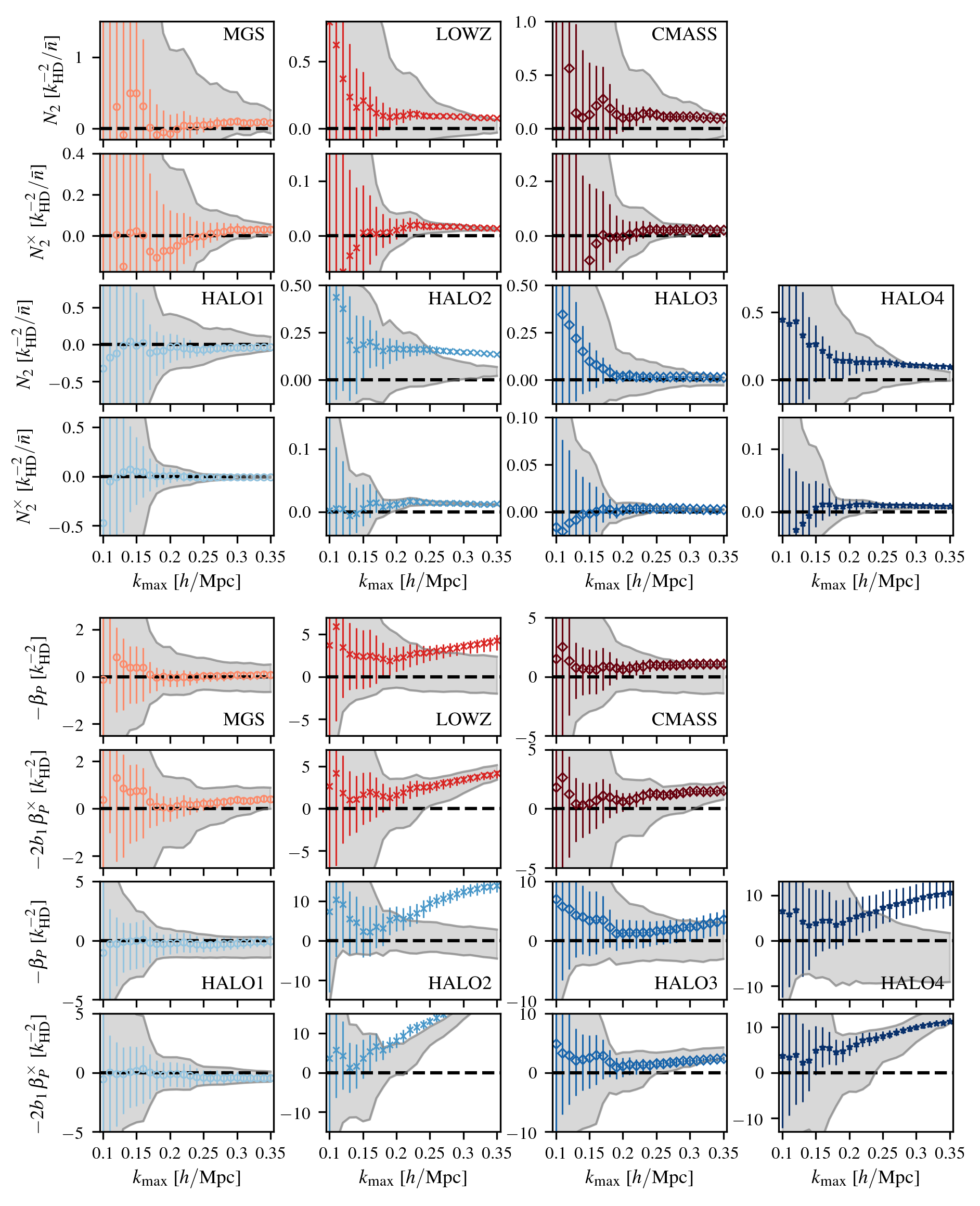}
  \caption{Constraints on scale-dependent stochasticity in the auto and cross power spectra, $N_2$ and
    $N_2^{\times}$ (upper half), and on higher-derivative effects, $\beta_P$ and $\beta_P^{\times}$ (lower
    half). Grey shaded areas represent the 1-$\sigma$ uncertainties from the individual auto and cross power
    spectrum fits as a function of $k_{\mathrm{max}}$, whereas the data points stem from their joint
    analysis. All results shown derive from fits in which all model parameter are allowed to vary.}
  \label{fig:N2N2x_trueM}
\end{figure*}

To begin with we check the consistency between the results of our model fits and the (model-independent)
determination of the constant shot noise parameter $N_0$ from large scale data only (see
Section~\ref{sec:measurements_bias_noise}). This is shown in Fig.~\ref{fig:N0_trueM}, where we plot the fully
marginalized 1-$\sigma$ uncertainties from the higher-derivative and scale-dependent noise models as a function
of $k_{\mathrm{max}}$ (red and blue bands), obtained from the joint $P_{gg}$ and $P_{gm}$ fits with all
parameters left free to vary. When compared to the large-scale measurement, depicted by the hatched band, we
find in general good agreement with the model predictions. However, for scales
$k_{\mathrm{max}} \gtrsim 0.25\,\iMpc$ the higher-derivative model develops a dependence on $k_{\mathrm{max}}$
that leads to a slight over-estimation of $N_0$ for all samples apart from HALO3. This is not the case when
allowing for scale-dependent stochasticity instead, and we only identify a trend with $k_{\mathrm{max}}$ for
HALO2, but on scales where Fig.~\ref{fig:fcf_Pgg+Pgm_trueM} already suggests a breakdown of the model.

Next, we consider the more interesting question whether we can put constraints on the scale-dependent noise or
higher-derivative parameters that enter the auto and cross power spectra. Their 1-$\sigma$ uncertainties,
plotted against $k_{\mathrm{max}}$, are shown in Fig.~\ref{fig:N2N2x_trueM}, where we have again chosen the most
conservative case in which none of the model parameters are held fixed. Each panel in the top half of the figure
corresponds to either $N_2$ or $N_2^{\times}$ for a given sample, and similarly for $\beta_P$ and
$2 b_1\beta_P^{\times}$ in the bottom half. This choice of variables for $\beta$'s is convenient since we are using the nonlinear matter spectrum to do the fits, in which case $P_\sigma=c_1=0$, thus one should recover $\beta_P=2 b_1\beta_P^{\times}$, see Eqs.~(\ref{eq:theory.Pgmk2}-\ref{eq:theory.Pggk2}). 
The results derived from the individual $P_{gg}$ and $P_{gm}$ fits are
shown as the gray shaded error bands, while the data points were obtained when jointly analyzing the two
observables. Note that the scale-dependent noise and higher-derivative parameters have units of $(\Mpc)^5$ and
$(\Mpc)^2$, respectively, so we show them as the dimensionless numbers that multiply the factors
$k_{\mathrm{HD}}^{-2}/\bar{n}$ or $k_{\mathrm{HD}}^{-2}$ with the (arbitrary) normalization scale
$k_{\mathrm{HD}} = 0.4\,\iMpc$.

\begin{figure}
  \centering
  \includegraphics[width=\columnwidth]{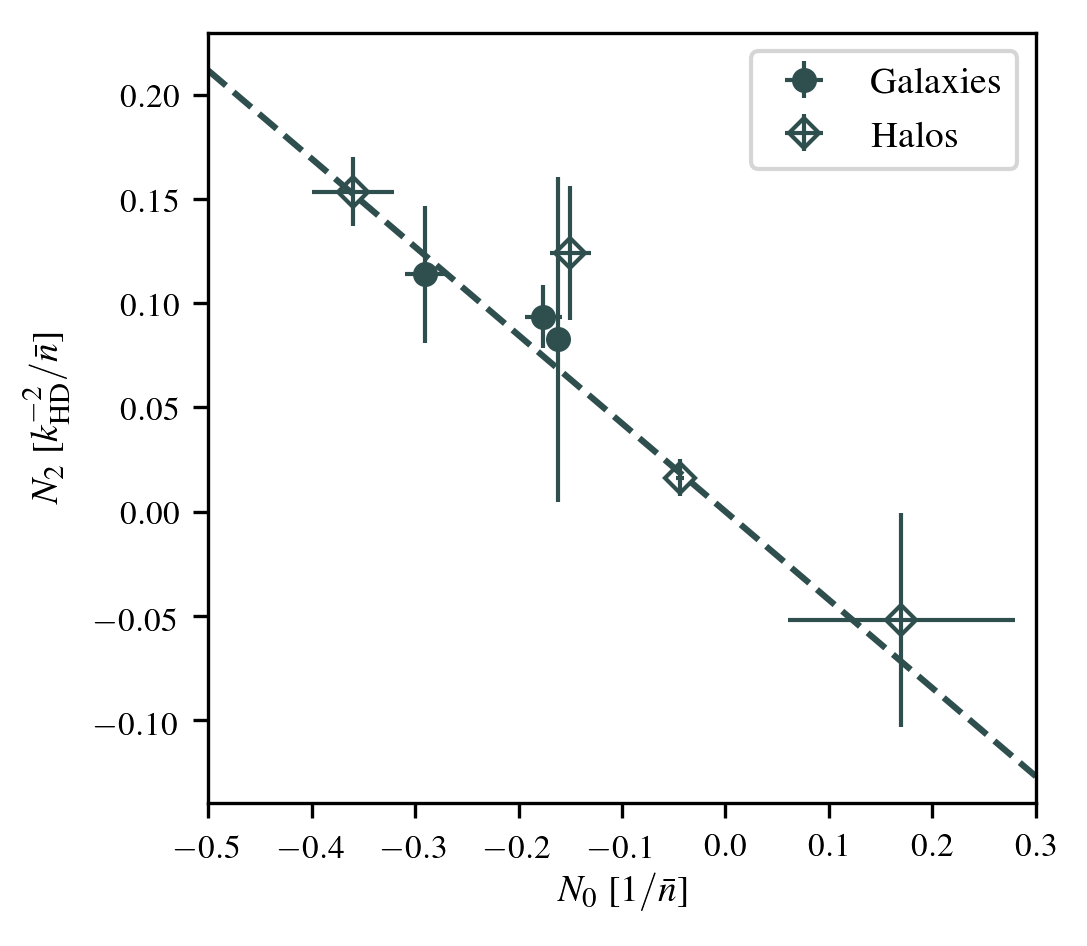}
  \caption{Measurement of the $N_2$ parameter from the joint auto and cross power spectrum fit at
    $k_{\mathrm{max}} = 0.27\,\iMpc$, plotted against the fiducial value for $N_0$ (see
    Table~\ref{tab:tracers}). The dashed line is a linear fit to this data with zero intercept.}
  \label{fig:N0vsN2}
\end{figure}

First we note that the uncertainties on $N_2$ from the auto power spectrum alone are rather large, so that
inclusion of even the smallest scales considered in our analysis does not yield a clear detection. The same is
also true for $\beta_P$ in the higher-derivative model. On the other hand, the cross power spectrum, which does
not contain $N_0$ and therefore has one free parameter less, gives non-zero values for $N_2^{\times}$ for all
samples except MGS and HALO1 at a significance above the 1-$\sigma$ level for $k_{\mathrm{max}} =
0.27\,\iMpc$. For the same samples we also find detections of $\beta_P^{\times}$, but their posterior means vary
strongly with increasing $k_{\mathrm{max}}$, while those of $N_2^{\times}$ stay mostly constant beyond a certain
scale.

For both models the parameter uncertainties shrink significantly once the auto and cross power spectrum are
combined, which comes as no surprise given our conclusion in Section~\ref{sec:consistency-ax} that the
stochasticity or higher-derivative terms are required to restore consistency between the two statistics. Again
it is noteworthy that the constraints on $N_2$ and $N_2^{\times}$ are much less sensitive to $k_{\mathrm{max}}$
than those for $\beta_P$ and $\beta_P^{\times}$, and additionally, they are in better agreement with the results
from the individual fits, as is particularly evident for LOWZ, HALO2 and HALO4. This lends further support to
the claim that scale-dependent stochasticity is the favored model extension for the samples under consideration.

Studying the recovered stochasticity constraints in more detail, we find that the $N_2$ parameter is
consistently a factor of a few larger than $N_2^{\times}$, meaning it is a stronger effect in the auto power
spectrum than it is in the cross spectrum. Indeed, we obtain a similar outcome when we evaluate the relative
contributions to the total auto or cross power spectrum from all combined galaxy bias loop corrections on the
one hand and the scale-dependent noise term on the other\footnote{For this we have used the best-fit parameter
  values obtained at $k_{\mathrm{max}} = 0.27\,\iMpc$, though the precise $k_{\mathrm{max}}$ value is
  irrelevant.}: while the two effects become nearly equal for $P_{gg}$ at $k \sim 0.3\,\iMpc$, even on such small
scales the stochastic term remains subdominant by at least an order of magnitude compared to one-loop bias
contributions in $P_{gm}$, which is consistent across all samples. This seems reasonable if the stochasticity in
the auto power spectrum is dominated by halo exclusion, which does not affect the cross power spectrum.

The halo exclusion effect should give rise to an additional signature in our constraints of $N_2$. As discussed
in Section~\ref{sec:stochasticity} we expect the stochasticity to approach the Poisson limit when $k$ becomes
large, which implies that for sub-Poissonian samples we should have $N_2 < 0$ and vice versa, $N_2 > 0$ for
super-Poissonian populations of galaxies or halos. Among our samples only HALO1 has super-Poissonian shot noise
on large scales and this is also the only case where we recover negative values of $N_2$. Moreover, we find a
strong (anti-) correlation between $N_2$ and the scale-independent shot noise parameter, which is shown in
Fig.~\ref{fig:N0vsN2} where we plot the fiducial values of $N_0$ from Section~\ref{sec:measurements_bias_noise}
versus the measured $N_2$ from the combination of $P_{gg}$ and $P_{gm}$ at $k_{\mathrm{max}} = 0.27\,\iMpc$. The
data allows us to perform a simple linear, one-parameter fit, which gives
\begin{equation}
  N_2(N_0) = \Big(-0.42 \pm 0.03\Big)\,\frac{N_0}{k_{\mathrm{HD}}^2}\,,
\end{equation}
and is shown by the dashed line. This suggests that the stronger the deviations from Poisson shot noise on large
scales, the bigger will be the response from scale-dependent stochastic contributions. The one outlier above the
dashed line (and not included in the fit) corresponds to HALO4, the most extreme biased tracer in our sample. 

Even though the behavior of the measured $N_2$ and $N_2^{\times}$ parameters seems reasonable in the context of
scale-dependent stochasticity, it is difficult to ascertain that this is the correct interpretation. That is
because its effect cannot be clearly distinguished from other higher-derivative terms such as
$\nabla^4\,\delta$. This term would give rise to a power spectrum contribution scaling as $k^4\,P_L(k)$, which
can appear identical to scale-dependent noise, as $P_L(k) \sim 1/k^2$ in the weakly nonlinear regime. For that
reason we have repeated our analysis for the auto power spectrum where we implemented the exact $k^4\,P_L(k)$
term in exchange for scale-dependent noise. We find indeed similar results that, however, display slightly
larger $\chi^2$ values consistently over all samples. Moreover, provided that the higher-derivative terms are a
valid perturbative expansion in powers of $(k\,R)^2$, higher order terms should be expected to become
increasingly relevant at larger $k$, so it seems peculiar why the second one should dominate, while the first is
mostly irrelevant. For these reasons we consider scale-dependent stochasticity as the more likely
explanation.  On the other hand, we caution that noise properties of halos/galaxies in our simulations may be unrealistic, in the sense that using a friends-of-friends halo finder  imposes strong exclusion properties that may not be realized in nature. See e.g.~\cite{GarRoz1911} for the impact of halo finder in the small-scale clustering properties of halos.

\begin{figure*}
  \centering
  \includegraphics[width=\textwidth]{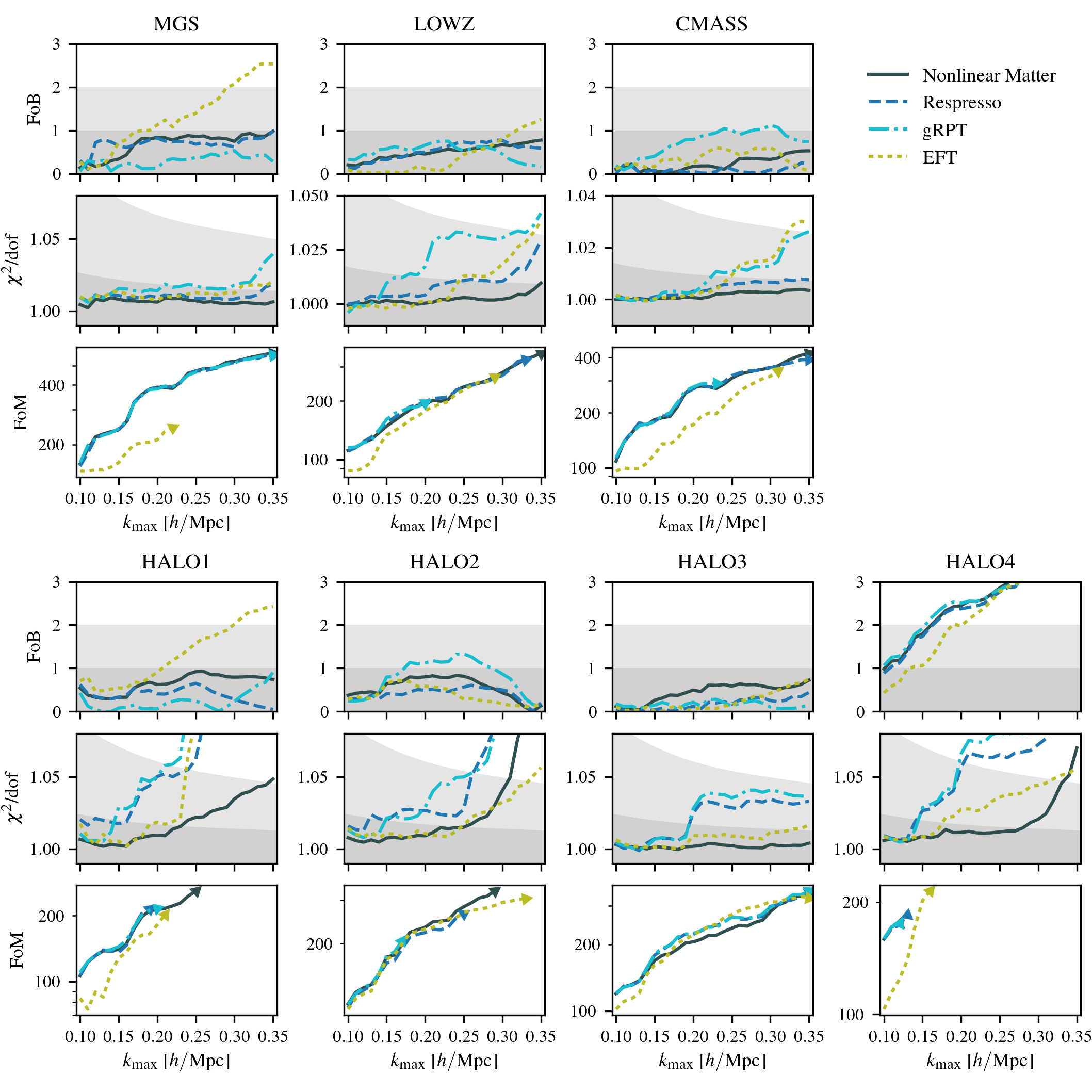}
  \caption{Same as Fig.~\ref{fig:fcf_Pgg+Pgm_trueM}, but only for the scale-dependent noise case with $\gamma_2$
    fixed to the excursion set relation, identified in Section~\ref{sec:results_true_matter} as the bias model giving best results over a
large range of scales when using the measured nonlinear matter power spectrum. Differently colored lines correspond to different choices of the
    nonlinear matter power spectrum, while the black solid line represents the previous result obtained with the
    measured nonlinear matter power spectrum.}
  \label{fig:fob_chi2_fom_matter_comparison}
\end{figure*}

\section{Results for surrogate matter models}
\label{sec:results_models}

In this section we compare the results obtained using the measured nonlinear matter power spectrum with the
various models presented in Section~\ref{sec:matter}: the two fully perturbative predictions from gRPT and EFT, as
well as the hybrid approach \texttt{RESPRESSO}. The aim of this analysis is to reveal how each of the three
options affects the range of validity and the FoM.

In order to simplify this comparison we only focus on the case that includes scale-dependent noise and
constrains $\gamma_2$ to follow the excursion set relation, which we previously identified as giving consistent
results over a large range of scales, independent of the particular sample under consideration. Furthermore, we
concentrate on the combination of the auto and cross power spectrum, as this allows for the most stringent test
of the matter modeling. While gRPT and \texttt{RESPRESSO} do not contain any additional model parameters, we
include the counter-term $c_1$ when performing fits using the EFT model and adopt a wide and flat prior in the
range $[-50,\,50]\,k_{\mathrm{HD}}^{-2}$ for this extra parameter. Repeating all previous steps for analyzing
the Markov chains, we arrive at the FoB, reduced $\chi^2$ and FoM as a function of $k_{\mathrm{max}}$ shown in
Fig.~\ref{fig:fob_chi2_fom_matter_comparison}, where the black, solid lines are for the true nonlinear matter
power spectrum and the various colored ones correspond to its three surrogate models.

As expected, we first note that modeling the matter power spectrum introduces a further source of inaccuracy,
which leads to a degradation of the reduced $\chi^2$ for all samples. However, the decrease in the
goodness-of-fit differs between the various models and we find that the EFT model is closer to the results obtained for the true nonlinear matter power for the halo samples, whereas \texttt{RESPRESSO} gives the best $\chi^2$ behavior for the galaxies, but curiously it is
somewhat worse for the two halo samples at redshift one, whereas gRPT has the largest $\chi^2$ values for
almost all cases and scales. These results are probably expected given the discussion in
Section~\ref{sec:matter} --- EFT has the advantage of an extra free parameter compared to gRPT and \texttt{RESPRESSO}, while the latter has some information from simulations already built in. One could improve the $\chi^2$ behavior of gRPT by including stress tensor corrections. In connection to this, \cite{SanScoCro1701} argued that having $\gamma_{21}$ free in the bias model partially compensates for the lack of stress tensor correction (as the two are exactly degenerate in the low-$k$ limit), but our results indicate that this is probably not enough because these terms are important at nonlinear scales where the shape of the $\gamma_{21}$ contributions is not fully degenerate with $k^2 P(k)$. Note also that there is mild evidence of overfitting caused by the extra parameter in the EFT, as it  leads to better  $\chi^2$  than using the actual nonlinear spectrum measured in the simulations for the LOWZ and HALO2 samples, but overall this does not appear as a strong concern. 

One may wonder to what extent are two-loop corrections in the matter field at play here. If this were the case, \texttt{RESPRESSO}  should be uniformly the  clear winner, particularly at low redshift (MGS, HALO1, HALO2) since it is the only method incorporating two-loop information. However, as seen in Fig.~\ref{fig:fob_chi2_fom_matter_comparison}, the situation is not as clear cut, in particular for the halos. Another issue at play is that  \texttt{RESPRESSO}  uses perturbation theory to compute the \emph{difference} in the power spectrum with respect to the Planck 2015 reference cosmology. The CMASS sample based on the \textsc{Minerva} simulations is the only sample whose cosmology is close to the reference cosmology, and we see that \texttt{RESPRESSO} performs clearly the best in that case, as expected. For the other samples, based on the  \textsc{LasDamas} simulations, the cosmology is fairly different and this might be playing a role, since  \texttt{RESPRESSO} assumes small deviations between target and reference cosmology. 

The FoB of the surrogate models is generally similar to the true matter power with the two notable exceptions of
MGS and HALO1, which become biased for the EFT model. In some other cases their FoB can also fall below that of
the true matter power, which should not be interpreted as the surrogates being more accurate, but rather that
the determination of the FoB is subject to some degree of noise. However, in total this means that the range of
validity determined from our combined criterion following Eq.~(\ref{eq:res1.kfail}) is dominated by the increase
in the reduced $\chi^2$. From the FoM panels of Fig.~\ref{fig:fob_chi2_fom_matter_comparison} (which terminate
at the breakdown scale $k_{\dagger}$) we see that \texttt{RESPRESSO} typically has the largest range of validity
across all samples, closely followed by the EFT. Apart from the HALO1 sample where all models fail at
$k_{\dagger} \sim 0.2\,\iMpc$, only gRPT suffers more considerable reductions, in particular for LOWZ, CMASS and
HALO2 with breakdown scales of the order $k_{\dagger} \sim 0.2 - 0.25\,\iMpc$, compared to $0.3\,\iMpc$ and
beyond for the true nonlinear matter power spectrum.

\begin{figure*}
  \centering
  \includegraphics[width=\textwidth]{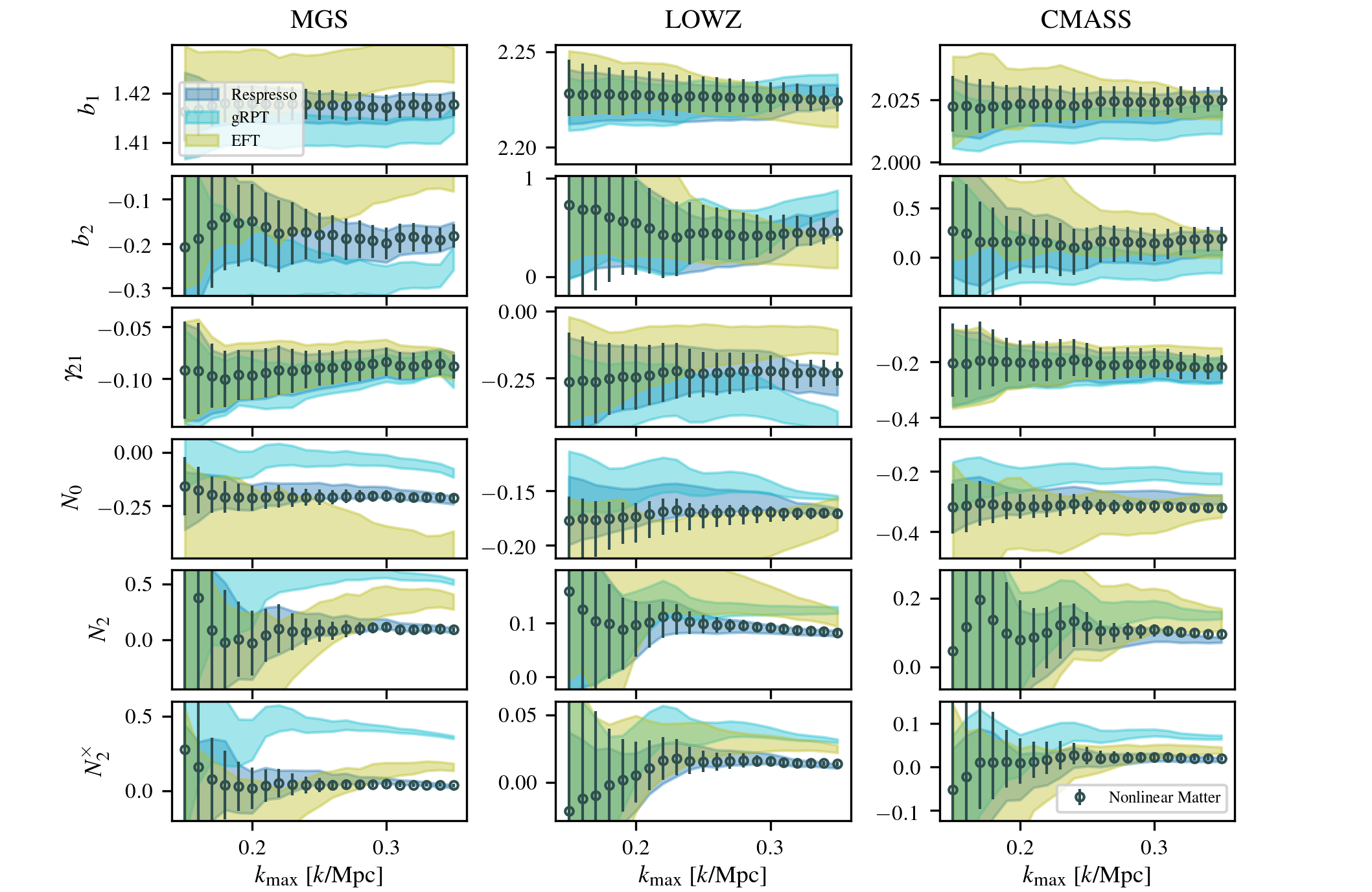}
  \caption{Constraints on the full set of model parameters (excluding the counter-term in case of EFT) as a
    function of $k_{\mathrm{max}}$ for various surrogate matter models (colored error bands) and the true
    nonlinear matter power spectrum (data points). The particular case considered here includes scale-dependent
    noise and assumes the excursion set relation for $\gamma_2$.}
  \label{fig:constraints_comparison}
\end{figure*}

Finally, the FoM is not affected at all when employing the \texttt{RESPRESSO} and gRPT models, but not
surprisingly decreases for the EFT due to its additional free counter-term parameter. Interestingly, although
its FoM can be reduced by up to $40\,\%$ at low $k_{\mathrm{max}}$, it is able to compensate  for the most part 
within its range of validity for all samples except MGS. This is in contrast with the modeling of the
matter power spectrum alone, where it was shown in \cite{OsaNisBer1903} that the FoM\footnote{However, we note
  that their definition of FoM was based on a set of cosmological parameters, which may correlate differently with 
  the counter-term than in our case, leading to a different impact on the FoM.} derived from \texttt{RESPRESSO} was superior to the EFT.

A breakdown of the matter modeling can also be disguised by other model parameters absorbing the emerging
differences. It is therefore interesting to check for any inconsistencies in the full parameter space compared
to the true matter power spectrum results, which we present for the galaxy samples in
Fig.~\ref{fig:constraints_comparison} using again the scale-dependent noise case with fixed $\gamma_2$ as an
example. Each panel shows the 1-$\sigma$ uncertainties at a given value of $k_{\mathrm{max}}$ for the three
models under consideration (colored bands), while the results using the true nonlinear matter power spectrum are
depicted by the black data points. In general, the agreement is good for the majority of the parameter space and
values of $k_{\mathrm{max}}$, but we see that gRPT tends to overestimate the large-scale shot noise amplitude
$N_0$ and both, gRPT and EFT, consistently return larger values of the scale-dependent noise parameters $N_2$
and $N_2^{\times}$. The latter is particularly evident for MGS, where the constraints suggest that the gRPT
model compensates for a lack of matter power on small scales. Further, albeit less consistent deviations occur
for $b_2$ and $\gamma_{21}$ in case of gRPT and EFT. To the contrary, \texttt{RESPRESSO} is an excellent match to
the results from the true matter power spectrum for all samples and scales. In principle one could include two-loop information in the other calculations to make the comparison with  \texttt{RESPRESSO} on a more equal footing, however this requires the introduction of further counter-terms (due to the increased sensitivity of the two-loop integrals to nonlinearities) and this probably will lead to a significant decrease in the overall FoM.

Finally, Fig.~\ref{fig:constraints_comparison} also demonstrates nicely that for the measured matter power
spectrum and, at least, \texttt{RESPRESSO} the posterior averaged mean parameter values are very insensitive to
the fitted range of scales, when their uncertainties decrease. This implies that the bias model is not
attempting to compensate for any unaccounted contributions and therefore is a further convincing point in favor
of the robustness of the chosen model.

\vspace*{1em}
\section{Conclusions}
\label{sec:conclusions}

This paper has addressed two leading questions: on which scales can one-loop perturbative models of galaxy bias
accurately describe measurements of two-point statistics, and how much freedom in terms of unknown bias
parameters do we need to allow for. In order to draw conclusions that are as general as possible, we have
systematically analyzed a diverse collection of tracers, comprising three galaxy and four halo samples at
different redshifts, each with statistical uncertainties corresponding to an effective volume of
$6\,(\Gpc)^3$. To robustly test one-loop galaxy bias, we used the measured nonlinear matter spectrum as this allows us to concentrate on bias independently of any issues related to the nonlinear evolution of matter description. In addition, by ignoring redshift-space distortions we are avoiding extra parameters that can mask failures of the bias model. 

Our most stringent test is based on MCMC fits to the auto power spectrum of galaxies (or halos) and their cross
spectrum with the underlying matter field. We tested various modeling assumptions, and quantitatively assessed the
model's performances by means of three metrics. Two of those, the figure of bias (FoB) and goodness-of-fit,
jointly determine the range of validity by guaranteeing an unbiased recovery of model parameters (here measured
in terms of the linear bias parameter $b_1$, which can be thought of as a proxy for the amplitude of the matter power spectrum) and a good match to the data. The third metric, the figure of merit
(FoM), quantifies the statistical uncertainty on model parameters (here also derived from $b_1$) and allows us
to identify a potential compromise between a reduced parameter set (i.e., more constraining power) and range of
validity.

The ``standard'' galaxy bias model for two-point statistics up to one-loop order contains five parameters: apart
from $b_1$, it depends on the quadratic and tidal biases, $b_2$ and $\gamma_2$, a contribution to tides due to nonlocal
gravitational evolution that appears at third order, $\gamma_{21}$, as well as a constant stochastic term $N_0$. In addition we have allowed
for either higher-derivative or scale-dependent stochasticity effects, and tested for the impact of fixing
$\gamma_2$ and $\gamma_{21}$ using excursion set and co-evolution relations. Our main findings from studying
these various options can be summarized as follows:
\begin{enumerate}[label=(\roman*)]
\item The standard five-parameter model applied to the auto power spectrum performs remarkably well and is
  applicable to the full tested range of scales, $k_{\mathrm{max}} = 0.35\,\iMpc$, for all our tracers except
  the two massive halo samples and LOWZ (the galaxy sample hosted by the most massive halos) in which case it
  fails at $k_{\mathrm{max}} \sim 0.2\,\iMpc$. In a joint analysis with the cross power spectrum the model
  develops inconsistencies in the nonlinear regime (as demonstrated by a quickly deteriorating goodness-of-fit),
  which lead to maximum validity scales that are reduced by up to $30\,\%$ with respect to the auto power
  spectrum alone.
\item The diminished model reliability for massive halos and the inconsistencies between auto and cross power
  spectra cannot be sufficiently resolved by the inclusion of the leading higher-derivative corrections. On the
  other hand, accounting for halo-halo exclusion through scale-dependent stochasticity brings significant
  improvements for all samples, allowing us to fit the measurements (even in combination with the cross
  spectrum) nearly up to $k_{\mathrm{max}} = 0.35\,\iMpc$ and likely beyond for some of the samples. From this
  we conclude that scale-dependent stochasticity has a stronger impact on galaxy/halo clustering than
  large-scale higher-derivative effects.
\item This conclusion is further supported by our constraints on the scale-dependent stochasticity and
  higher-derivative parameters. From the joint analysis at $k_{\mathrm{max}} = 0.25\,\iMpc$ and for the most
  conservative case (all model parameters are varied in the Markov chain), we find detections of the former at
  the level of $1.4$ to $7$-$\sigma$ in the auto power spectrum (only our MGS sample is below the $1$-$\sigma$
  threshold) and similar, but slightly less significant results in the cross spectrum. These detections become
  even more pronounced at larger wave numbers, while the mean parameter values are largely independent of
  $k_{\mathrm{max}}$. We also find clear detections of the higher-derivative parameters, but in contrast these
  depend strongly on the range of scales included in the fit. We further find that scale-dependent
  stochasticity affects the auto power spectrum more substantially than the cross spectrum and a tight
  correlation between the scale-independent and -dependent parameters (see Fig.~\ref{fig:N0vsN2}), both of which are in line with our
  interpretation in terms of halo-halo exclusion.
\item Application of the excursion set relation for $\gamma_2$ (see Fig.~\ref{fig:g2ECS} of how this compares
  with the local Lagrangian relation and precise measurements from \cite{LazSch1712}) improves the FoM without
  diminishing the validity ranges for all samples and models (the only notable exception being the high-mass
  halos at redshift $z=1$, the most biased tracer in our sample), which makes this a preferred choice for reducing the parameter space. We find that
  $\gamma_{21}$ is generally in good agreement with its co-evolution relation (and constrained to be non-zero, particularly when $\gamma_2$ is fixed) and when used in combination with the excursion set relation for $\gamma_2$ can give rise
  to an even more substantial enhancement of the FoM. Therefore, it typically provides the best compromise
  between constraining power and scales on which the model is applicable. However, we caution that in this case
  the latter can vary  from sample to sample, so great care should be taken when choosing to apply this
  approximation to real survey data.
\item Combination of all previous points suggests that the standard model with tidal bias fixed by the excursion
  set relation (four free parameters in total) provides a robust modeling choice for the auto power spectrum of
  the less massive halos in our set of samples and galaxy populations living in those (MGS, CMASS). For the more
  massive halos and the LOWZ galaxy sample it is most beneficial to include an extra parameter corresponding to
  scale-dependent stochasticity. This is also the preferred option when considering combinations of the auto and
  cross power spectrum, which might be relevant in joint studies of galaxy clustering and weak lensing. In this case
  there would be six free bias parameters to account for independent stochastic contributions in the two
  spectra.
\end{enumerate}
  
All of these results (Section~\ref{sec:results_true_matter}) were derived using the measured power spectrum of the
underlying matter field, which enters in the models of the galaxy auto and cross power spectra through the linear bias terms (see
Eqs.~\ref{eq:theory.Pgm_1loop} and \ref{eq:theory.Pgg_1loop}). This was done explicitly to test one-loop galaxy bias independently of one-loop corrections to the matter field.

In Section~\ref{sec:results_models}, we tested the impact of modeling the nonlinear matter spectrum for the
complete set of biased tracers, using the ``best choice'' of bias priors, i.e.  including scale-dependent noise
and fixing the quadratic tidal tensor bias $\gamma_2$ to follow the excursion set relation, which as discussed
above was identified as giving consistent results over a large range of scales. We considered two fully
perturbative surrogates of the matter power spectrum, gRPT and EFT, as well as the hybrid approach
\texttt{RESPRESSO}. While \texttt{RESPRESSO} contains no free parameters, the EFT model includes one free
parameter associated to a counter-term arising from stress-tensor corrections, which is effectively built-in
in~\texttt{RESPRESSO}.  In principle, gRPT should also include stress tensor corrections, but we left these out
following the implementation in~\cite{SanScoCro1701}, which argued that having $\gamma_{21}$ free in the bias
model partially compensates for this choice (as the two are exactly degenerate in the low-$k$ limit). Compared
to the results with the measured matter power spectrum all three models have a similar performance, but in some
cases reduced validity ranges. We find that the best surrogate (in terms of validity and similarity of the
recovered mean posterior values) is \texttt{RESPRESSO}, followed by the EFT (see
Fig.~\ref{fig:constraints_comparison}). Due to its extra parameter the latter can lag behind in FoM, but can
mostly compensate for this at larger $k_{\mathrm{max}}$. \correction{Since the assumptions on galaxy bias were
  held fixed, these results also give an indication for how well each of the three models are able to describe
  the measured matter power spectrum. The conclusions one would draw based on this alone are consistent with
  previous studies (e.g. \cite{OsaNisBer1903,SanScoCro1701}).}

As stated above, all MCMC fits carried out in this analysis made use of statistical measurement uncertainties
corresponding to an effective volume of $6\,(\Gpc)^3$. This is significantly below the total volume that will be
observed by upcoming galaxy surveys such as DESI and Euclid, but one should keep in mind that for the actual
clustering analyses the total volume will be split into a number of redshift slices, for which our adopted
volume here should be more than representative. Nonetheless, we have investigated how our derived model validity
ranges scale with a more restrictive breakdown criterion, which we showed can to first order be interpreted as
an increase in the effective volume. While the breakdown scales move towards lower wave numbers, this test
revealed that they do so more slowly for the model that includes scale-dependent stochasticity. For that reason
its maximum FoM values become level or even better than those of the standard model, suggesting that it might be
the optimal modeling choice for any of our samples at larger volumes.

Although our collection of tracers span a large variety, we miss samples consisting of halos less massive than
$10^{13}\,M_{\odot}$, as well as galaxies that match the clustering properties of emission line galaxies, which
are the main targets of the DESI and Euclid surveys. There is no guarantee that our conclusions regarding the
excursion set and co-evolution relations for $\gamma_2$ and $\gamma_{21}$, respectively, would still be valid
for these types of galaxies. In general, galaxies are expected to inherit their bias from their host halos and a
deciding factor in the similarity between the two is the satellite fraction of the galaxy population. For all
three of our mock galaxy samples the satellite fraction is rather low ($\simeq 10\%$), making them reliable tracers of the halo centers. As these obey the excursion set
relation for the tidal bias, this likely explains why we have also obtained good results for our galaxy
samples. On the other hand, for populations with higher satellite fractions one could consider using a given HOD
model to reweight the tidal bias of the host halos according to the expected central and satellite fractions. We
leave a more detailed exploration of this possibility for future work.

In forthcoming publications we will extend the results in this paper to account a variation of cosmological parameters~\cite{Pezzotta}, as well as extending the test of one-loop galaxy bias to the one-loop bispectrum~\cite{Eggemeier}.

\acknowledgements

We thank T.~Lazeyras for making his measurements of the tidal bias parameters available to us, and R.~Sheth and
R.~Smith for useful discussions. AE acknowledges support from the European Research Council (grant number
ERC-StG-716532-PUNCA), while MC acknowledges support by the Spanish Ministry of Science MINECO under grant
PGC2018-102021. AGS acknowledges support by the Excellence Cluster ORIGINS, which is funded by the Deutsche
Forschungsgemeinschaft (DFG, German Research Foundation) under Germany’s Excellence Strategy - EXC-2094 -
390783311. This research made use of \texttt{matplotlib}, a Python library for publication quality graphics
\citep{Hunter:2007}.

This work was finalized during the Covid-19 outbreak. The authors would like to thank all essential workers
around the world that continue to make huge sacrifices to overcome this pandemic.

\bibliography{refs}

\end{document}